\documentclass[10pt,aps,prb,secnumarabic,nobibnotes,twocolumn,superscriptaddress,floatfix]{revtex4-2}
\usepackage{amsfonts}
\usepackage{mathrsfs}
\usepackage{amsmath}% needed for subequations
\usepackage{color}
\usepackage{natbib}
\usepackage{graphicx}
\usepackage{bm}% bold maths
\usepackage{amssymb}
\usepackage{xspace}
\usepackage{epstopdf}
\usepackage{dcolumn}
\usepackage{multirow}
\usepackage[colorlinks=true, pdfstartview=FitV, linkcolor=blue, citecolor=blue, urlcolor=blue]{hyperref}
\usepackage{wrapfig}
\usepackage{makecell}
\usepackage{placeins}

\makeatletter

\newcommand{\Rmnum}[1]{\expandafter\@slowromancap\romannumeral #1@}
\makeatother

\begin{document}
\title{Second-order Real Nodal Lines in Nodal Surface Semimetals}

\author{Yunye Wang}
\affiliation{Shenzhen Research Institute of Shandong University, Shenzhen 518057, China}

\author{Ruopu Zhao}
\affiliation{Shenzhen Research Institute of Shandong University, Shenzhen 518057, China}

\author{Tianqi Zhang}
\affiliation{Shenzhen Research Institute of Shandong University, Shenzhen 518057, China}

\author{Weikang Wu}
\email{weikang\_wu@sdu.edu.cn}
\affiliation{Shenzhen Research Institute of Shandong University, Shenzhen 518057, China}

\begin{abstract}
Real nodal lines (RNLs) featuring real Chern numbers and second-order boundary modes have attracted widespread attention. In all previously reported realizations, an RNL is linked by another nodal line, and whether an RNL can be linked by other types of band degeneracies has remained open.
Here, we propose a second-order real nodal-line semimetal in which a pair of RNLs is linked by a nodal surface. We show that this state can be realized in spinless systems with both $\mathcal{PT}$ and nonsymmorphic $S_{2z}\mathcal{T}$ symmetries, where the $S_{2z}\mathcal{T}$-enforced nodal surface prevents the pair of RNLs from annihilation. Each nodal line carries a nontrivial real Chern number $\nu_R=1$, giving rise to topological hinge Fermi arcs located at a pair of $\mathcal{PT}$-related hinges. Guided by this construction, we identify the interpenetrated graphene network (IGN) as a promising material realization. First-principles calculations confirm that a pair of nodal lines traversing the Brillouin zone are linked by a nodal surface and each nodal line carries double nontrivial $\mathbb{Z}_2$ charges. The bulk-boundary correspondence of IGN manifests as a pair of hinge Fermi arcs together with drumhead surface states. Our work establishes nodal surfaces as a new linking partner for real nodal lines and provides a roadmap for exploring higher-order real topology in carbon-based and other light-element systems.
\end{abstract}
\maketitle

\emph{\color{blue}Introduction. --- }
Topological metals and semimetals have been attracting tremendous research interest over the past decade, owing to their exotic low-energy band structures with topologically protected nodal degeneracies~\cite{Bansil2016Colloquium-RMP,Burkov2016Topological-NM,Yang2016Dirac-SPIN,Armitage2018Weyl-RMP,Yu2022Encyclopedia-SB}. The band degeneracies protected by crystalline symmetries or topological invariants appear as zero-dimensional nodal points~\cite{Wan2011Topological-PRB,Xu2011Chern-PRL,Fang2012Multi-PRL,Young2012Dirac-PRL,Wang2012Dirac-PRB,Wang2013Three-PRB}, one-dimensional nodal lines~\cite{Weng2015Topological-PRB,Chen2015Nanostructured-NL,Kim2015Dirac-PRL,Yu2015Topological-PRL,Fang2016Topological-CPB,Yu2019Quadratic-PRB}, or even two-dimensional nodal surfaces~\cite{Liang2016Nodal-PRB,Zhong2016Towards-N,Wu2018Nodal-PRB}. Among these, nodal lines are especially intriguing, because they form extended manifolds in momentum space and enable a rich variety of topological structures, including single lines~\cite{Chen2015Nanostructured-NL,Kim2015Dirac-PRL}, chains~\cite{Bzdusek2016Nodal-N,Yu2017From-PRL,Wang2017Hourglass-NC}, links~\cite{Chang2017Topological-PRL,Yan2017Nodal-PRB}, or even knots~\cite{Bi2017Nodal-PRB}.

Nodal-line semimetals can arise in systems with either spatial inversion $\mathcal{P}$ and time-reversal $\mathcal{T}$ symmetries, or mirror symmetry~\cite{Fang2016Topological-CPB}. In particular, for $\mathcal{PT}$-symmetric spinless systems, a nodal line can be characterized by the $\mathbb{Z}_2$ Berry phase that is quantized due to $\mathcal{PT}$. Moreover, with negligible spin-orbit coupling (SOC), $(\mathcal{PT})^2=1$ ensures that both the Hamiltonian and the Bloch wave functions can be chosen real-valued~\cite{Morimoto2014Weyl-PRB,Fang2015Topological-PRB,Zhao2016Unified-PRL,Zhao2017PT-PRL}. This allows an additional $\mathbb{Z}_2$ topological quantity, the second Stiefel-Whitney number $w_2$, equivalently the real Chern number $\nu_R$~\cite{Zhao2017PT-PRL,Ahn2018Band-PRL,Ahn2019Stiefel-CPB,Wang2020Boundary-PRL,Chen2021Graphyne-PRB,Chen2022Second-PRL}. The nodal lines in such systems are therefore doubly charged. A conventional nodal line is protected solely by a $\pi$ Berry phase, whereas a real nodal line (RNL) simultaneously exhibits both the $\pi$ Berry phase and a nontrivial real Chern number.

Distinguished from the conventional nodal line, an RNL with nontrivial real Chern number exhibits distinct topological features. The RNL can only be created and annihilated in pairs, due to the nontrivial $\nu_R$~\cite{Ahn2019Stiefel-CPB}. Since the real Chern insulator is a second-order topological insulator hosting a corner state~\cite{Benalcazar2017Quantized-S,Song2017d-PRL,Langbehn2017Reflection-PRL,Schindler2018High-SA,Sheng2019Two-PRL,Chen2021Graphyne-PRB}, this indicates the second-order topology of RNLs, manifested in the hinge Fermi arcs located on a pair of hinges connected by $\mathcal{PT}$~\cite{Wang2020Boundary-PRL,Chen2022Second-PRL}. This distinctive band topology of RNLs has been predicted in a few electronic systems with negligible SOC~\cite{Wang2019Higher-PRL,Lee2020Graphdiyne-nQM,Chen2022Second-PRL}, phonon systems~\cite{Wang20243D-AFM,Wang2024Real-AM,Han2024Crossed-PRB}, and even artificial periodic systems~\cite{Xue2023Stiefel-NC,Xiang2024Demonstration-PRL,Ma2024Observation-PRL}. More recently, a general construction scheme based on stacking two-dimensional $\mathcal{PT}$-symmetric Dirac semimetals was developed, leading to abundant candidate materials including more than one hundred transition-metal dichalcogenides~\cite{Li2025General-PRB}.

\begin{figure}[t]
\centering
\setlength{\abovecaptionskip}{6pt}
\includegraphics[width=\columnwidth]{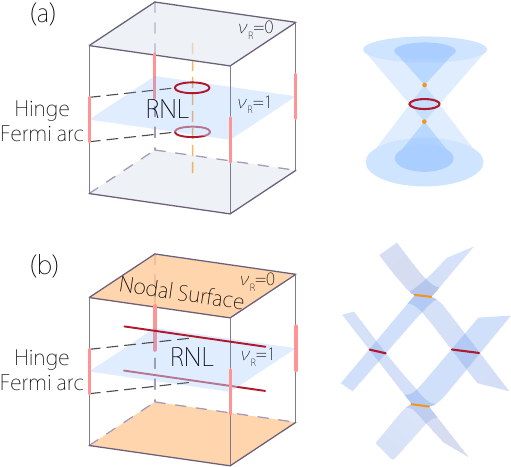}
\caption{\label{fig:idea}Schematic figures showing two types of linking configurations for RNLs. (a) Conventional RNL configuration, where a pair of real nodal loops formed by two middle bands (red line) is linked by the nodal lines formed by the bottom (top) bands (orange dashed line). The corresponding three-dimensional band dispersion is shown on the right. (b) The nodal-surface-linked RNL configuration proposed in this work. A pair of RNLs traversing the BZ (red line) is linked by a nodal surface formed by the bottom (top) band degeneracy (orange plane). In the BZ, the $\nu_R=1$ slices between the RNLs dictate hinge Fermi arcs (pink lines). The corresponding three-dimensional band dispersion is shown on the right. The nodal surface serves as a barrier to prevent the pair annihilation of RNLs.}
\end{figure}

Generally, the RNL formed by double band inversion is not alone but must be linked by an additional nodal line formed by the two bottom (top) bands [see Fig.~\ref{fig:idea}(a)]. For example, the ABC-stacked graphdiyne, proposed as the first material candidate, hosts RNLs linked by another straight nodal line formed between the two topmost occupied bands~\cite{Ahn2018Band-PRL,Chen2022Second-PRL}. Even in phonon systems where crossed real nodal lines were proposed in gold monobromide, the linking partner is still a nodal line rather than another type of degeneracy~\cite{Han2024Crossed-PRB}. To date, all reported RNLs' linking partner is always another nodal line. Whether an RNL can be linked by other types of band degeneracies remains an open question.

In this work, we propose a previously unrecognized type of RNL linked by a nodal surface. The linking structure is constructed in a spinless system with both $\mathcal{PT}$ and $S_{2z}\mathcal{T}$ symmetries, where $S_{2z}$ is the twofold screw axis. $S_{2z}\mathcal{T}$ dictates the essential existence of a nodal surface at the $k_z = \pi$ plane, which inhibits the annihilation of a pair of RNLs located on opposite sides of the nodal surface. The configuration is illustrated in Fig.~\ref{fig:idea}(b). To validate this scheme, we construct a generalized tight-binding model that illustrates the possibility of the nodal-surface-linked RNLs. We show that a pair of RNLs in the linking structure both possess nontrivial $\nu_R = 1$. The nontrivial topology in the bulk manifests as topological hinge modes located at a pair of hinges connected by $\mathcal{PT}$. Guided by these theoretical insights, we further identify the carbon allotrope, interpenetrated graphene network (IGN), as a promising material candidate. Combined with first-principles calculations, we find that a pair of nodal lines traversing the Brillouin zone are indeed linked by a nodal surface at $k_z = \pi$ in the electronic band structure of IGN. We evaluate real Chern numbers and boundary states for the nodal lines in IGN and identify the RNL state with a pair of hinge Fermi arcs. Furthermore, we demonstrate the robustness of the RNLs and their hinge modes against screw-symmetry breaking, which removes the nodal surface. This work opens the door to a new class of RNLs and may stimulate further studies on their realization in realistic material systems.

\emph{\color{blue} General idea. --- }
First, let us introduce the general idea for realizing RNLs in nodal surface semimetals. Consider a nodal-surface spinless system that is protected by the combined $S_{2z}\mathcal{T}$ symmetry, with $S_{2z}\equiv\{C_{2z}|00\frac{1}{2}\}$ being a twofold screw rotation and $\mathcal{T}$ the time-reversal symmetry. Without loss of generality, we choose the screw axis along the $z$ direction. $S_{2z}\mathcal{T}$ enforces the presence of a nodal surface at the $k_z = \pi$ plane, owing to $(S_{2z}\mathcal{T})^2 = -1$~\cite{Wu2018Nodal-PRB}. To realize a second-order real nodal line (SORNL), the system is required to preserve the spacetime inversion symmetry $\mathcal{PT}$ with $(\mathcal{PT})^2 = 1$. It ensures the Hamiltonian to be real, allowing for the definition of a real Chern number $\nu_R$. Here, we focus on the nonmagnetic system, i.e., $\mathcal{T}$ is preserved. In other words, we consider a centrosymmetric system.

In the implementation, we consider the nodal line traversing the whole Brillouin zone (BZ). Such a kind of nodal line is topologically distinct from a closed nodal ring, because it cannot be continuously deformed into a point. Mathematically, given that the three-dimensional BZ is topologically equivalent to a three-torus $\mathbb{T}^3$, a BZ-traversing nodal line on $\mathbb{T}^3$ can be classified under the fundamental homotopy group $\pi_1 (\mathbb{T}^3)$ and features a nontrivial $\mathbb{Z}^3$ index~\cite{Li2017Type-PRB}. With the $\mathcal{PT}$ symmetry, a nodal line is characterized by a quantized $\pi$ Berry phase for any closed path encircling the nodal line~\cite{Fang2016Topological-CPB}. Thus, there must be a pair of nodal lines which both carry a flux of $\pi$. Each of the two nodal lines cannot be annihilated by itself; they can only annihilate in a pair.

Next, we introduce the $S_{2z}\mathcal{T}$-enforced nodal surface at the $k_z = \pi$ plane. The pair of nodal lines lies on either side of the surface and is related by $\mathcal{P}$. The nodal surface is formed by the double degeneracy of the two bottom (top) bands. Together with the nodal lines, they form a linking configuration. One such scenario is shown in Fig.~\ref{fig:idea}(b). Obviously, the nodal surface acts as a barrier, preventing the nodal lines from merging and annihilating.

Apparently, the nodal lines embedded in a nodal surface are topologically distinct from the conventional nodal lines. The latter are characterized by a $\mathbb{Z}_2$ topological classification based on the quantized Berry phase. In contrast, besides the $\pi$ Berry phase, the nodal line linked by a nodal surface would also carry a $\mathbb{Z}_2$ monopole charge, just like its conventional counterpart linked by another nodal line. The $\mathbb{Z}_2$ monopole charge is characterized by the nontrivial real Chern number $\nu_R$; hence it is an RNL.

We have two remarks here. First, the RNLs linked by a nodal surface must traverse the entire Brillouin zone. Otherwise, each of them as a ring would be continuously deformed into a point and then be gapped out without meeting a singularity in the process, which contradicts the nontrivial $\mathbb{Z}_2$ monopole charge.

Second, although the nodal surface prevents the pair of RNLs from annihilating each other, they can still annihilate in a pair when moving towards the $k_z = 0$ plane. However, this does not contradict the fact that they are RNLs since each of them carries a nontrivial $\nu_R$.

\begin{figure}[t]
\centering
\includegraphics[width=\columnwidth]{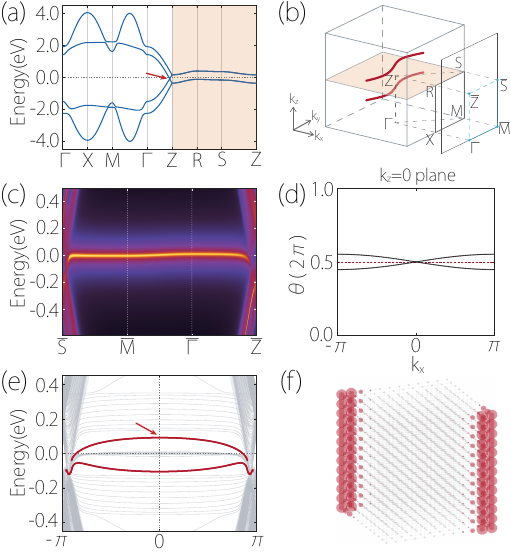}
\setlength{\abovecaptionskip}{6pt}
\caption{\label{fig:tb}The tight-binding model. (a) Typical bulk band structures. The RNL is labeled by the arrow, while the shaded area denotes the nodal surface at the $k_z=\pi$ plane. (b) The bulk BZ and the (100) surface BZ. The red lines indicate the pair of RNLs located on the two sides of a nodal surface at $k_z=\pi$ (orange plane). (c) The local density of states on the (100) surface, which shows drumhead-like states. (d) The Wilson-loop spectrum at $k_z=0$. The spectrum features a crossing at $\theta=\pi$, indicating the nontrivial $\nu_R$. (e) The energy spectrum for a nanorod sample. The nanorod has a $12\times12$-cell cross section with open boundaries along $x$ and $y$ directions. The periodicity is kept along the $z$ direction. The hinge bands are marked in red. (f) Spatial distribution of the $\mathcal{PT}$-related hinge arc mode that corresponds to the state marked by the red arrow in (e).}
\end{figure}

\emph{\color{blue}Tight-binding model. --- }
To validate our proposed mechanism, we construct a minimal four-band spinless tight-binding (TB) model constrained by space inversion $\mathcal{P}$ and twofold screw $S_{2z}$ symmetries, as well as the time reversal $\mathcal{T}$. 
The construction details are presented in the Supplementary Material~\cite{SupplementaryMaterial}.

A typical band structure of the TB model is given in Fig.~\ref{fig:tb}(a). One observes that two middle bands cross along the $\Gamma$-$Z$ path, while the two lower (upper) bands are degenerate along the BZ boundary ($Z$-$R$-$S$-$Z$), forming a nodal surface on the $k_z=\pi$ plane. A careful scan over the BZ reveals that the middle-band crossing forms a pair of nodal lines traversing the whole BZ and located symmetrically on the two sides of the $k_z=\pi$ plane, as plotted in Fig.~\ref{fig:tb}(b). This forms the linking configuration proposed in Fig.~\ref{fig:idea}(b).

To confirm the topological character, we evaluate the real Chern number $\nu_R$ for the two nodal lines. Specifically, we consider the horizontal momentum slices on the two sides of a nodal line. In general, $\nu_R$ can be directly read from the Wilson loop spectrum, by counting the number of crossings at $\theta=\pi$~\cite{Ahn2018Band-PRL,Chen2021Graphyne-PRB,Yue2024Stability-PRB}. If the crossing number is odd, the slice is topologically nontrivial with $\nu_R=1$; otherwise, it is trivial with $\nu_R=0$. The calculation details are provided in the Supplementary Material~\cite{SupplementaryMaterial}. As illustrated in Fig.~\ref{fig:tb}(d), the spectrum on the $k_z=0$ plane exhibits an odd number of protected crossings at $\theta=\pi$, confirming a nontrivial real Chern number $\nu_R=1$. However, for the $k_z=\pi$ plane, its Wilson-loop spectrum is topologically trivial, yielding $\nu_R=0$ [see Fig.~S4(b) in the Supplementary Material~\cite{SupplementaryMaterial}]. Since the nodal line is the only nodal degeneracy between $k_z=0$ and $k_z=\pi$ planes, the change of $\nu_R$ indicates that the nodal line carries a nontrivial real Chern number; hence, it is an RNL. The same analysis applies to the other nodal line related by $\mathcal{P}$, which also carries $\nu_R=1$.

Alternatively, for a centrosymmetric system, $\nu_R$ can be efficiently evaluated using the parity eigenvalues of the occupied bands at the time-reversal-invariant momenta (TRIMs)~\cite{Ahn2018Band-PRL,Chen2021Graphyne-PRB}. For a 2D plane covering four TRIMs in the Brillouin zone, $\nu_R$ is determined by
\begin{equation}
    (-1)^{\nu_R} = \prod_{\Gamma_i \in \text{TRIM}} (-1)^{\left\lfloor N_{-}(\Gamma_i)/2 \right\rfloor},
\end{equation}
where $\lfloor \dots \rfloor$ represents the floor function, and $N_-(\Gamma_i)$ is the number of occupied valence bands with negative parity eigenvalues at the TRIM point $\Gamma_i$. For the case of our TB model, it is convenient to do the evaluation for the $k_z=0$ and $k_z=\pi$ planes. 

\begin{table}[t]
\caption{Parity eigenvalues of the tight-binding model at the eight $\mathcal{P}$-invariant points. $N_-$ denotes the number of occupied bands with negative parity eigenvalues. The $k_z=0$ and $k_z=\pi$ planes have $\nu_R=1$ and $\nu_R=0$, respectively. \label{table:tb_model}}
\begin{ruledtabular}
\begin{tabular}{l cccc c cccc}
& \multicolumn{4}{c}{$k_z=0$} & & \multicolumn{4}{c}{$k_z=\pi$} \\[1pt]
\cline{2-5} \cline{7-10}
\rule{0pt}{10pt}& $\Gamma$ & $X$ & $M$ & $Y$ & & $Z$ & $R$ & $S$ & $T$ \\[1pt]
\hline
\rule{0pt}{9pt}$N_-$ & 2 & 1 & 1 & 0 & & 1 & 1 & 1 & 1 \\[1pt]
$\nu_R$ & \multicolumn{4}{c}{1} & & \multicolumn{4}{c}{0} \\
\end{tabular}
\end{ruledtabular}
\end{table}

The calculated values of $N_- (\Gamma_i)$ at all eight TRIM points are summarized in Table~\ref{table:tb_model}. In perfect agreement with the Wilson loop results, we find $\nu_R = 1$ for the $k_z = 0$ plane and $\nu_R = 0$ for the $k_z = \pi$ plane. The switch in topology with $k_z$ from $\nu_R = 1$ to $\nu_R = 0$ indicates that the nodal lines in between carry a nontrivial monopole charge, confirming that they are indeed RNLs.

In addition, we note that each nodal line also carries a 1D topological charge characterized by the quantized Berry phase~\cite{Fang2016Topological-CPB}:
\begin{equation}\label{eq:Berry_phase}
    w = \frac{1}{\pi} \oint_{C} \mathcal{A}(\bm{k}) \cdot d\bm{k} \mod 2,
\end{equation}
where $C$ is a closed path encircling the nodal line. The calculated Berry phase for a loop around each nodal line is $\pi$, corresponding to $w=1$. Hence, the RNLs in our TB model possess dual $\mathbb{Z}_2$-valued topological charges $(w,\nu_R)=(1,1)$. 

\begin{figure}[t]
\centering
\setlength{\abovecaptionskip}{6pt}
\includegraphics[width=\columnwidth]{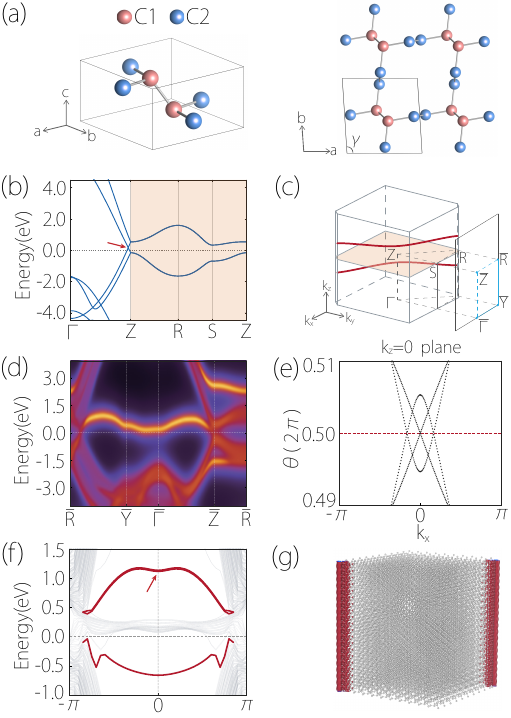}
\caption{\label{fig:ign} (a) Side view and top view of the crystal structure for IGN. C1 (red) and C2 (blue) denote $sp^3$- and $sp^2$-hybridized carbon, respectively. (b) Calculated electronic band structures of IGN. The RNL is labeled by the arrow, while the shaded area denotes the nodal surface at the $k_z=\pi$ plane. (c) The bulk BZ and the (010) surface BZ. The red lines indicate the pair of RNLs located on the two sides of a nodal surface at $k_z=\pi$ (orange plane). (d) The (010) surface spectrum with drumhead states. (e) The Wilson-loop spectrum for the $k_z=0$ plane. There are three crossings at $\theta=\pi$, indicating that $\nu_R=1$. (f) The energy spectrum for a nanorod sample with the $10\times10$-cell cross section. The periodicity is kept along the $z$ direction. The hinge bands are marked in red. (g) Spatial distribution of a $\mathcal{PT}$-related hinge arc mode at $k_z = 0$ [marked by the red arrow in (f)].}
\end{figure}

These nontrivial charges dictate the simultaneous emergence of first-order and second-order topological boundary states. In Fig.~\ref{fig:tb}(c), we show the energy spectrum for the (100) surface. The drumhead-like surface states are clearly observed, connecting to the surface projection points of RNLs. This is consistent with the nontrivial 1D topological charge $w=1$.

More crucially, the nontrivial real Chern number guarantees the existence of topological hinge states, whose position can be inferred from $\nu_R$ for the 2D slice in bulk BZ. As aforementioned, the $k_z = 0$ plane is nontrivial with $\nu_R=1$, and thus behaves as a 2D real Chern insulator which must host at least a pair of $\mathcal{PT}$ connected corner states. In fact, all the slices with $k_z \in (-k_b, k_b)$ (where $k_b$ is the lower bound of RNLs) between the two RNLs would have $\nu_R=1$, and feature corner modes. All these corner states constitute a hinge Fermi arc which terminates at the projections of RNLs. To verify this higher-order topology, we calculate the energy spectrum for a nanorod geometry with open boundary conditions along both the $x$ and $y$ directions shown in Fig.~\ref{fig:tb}(e). The structure is periodic along the $z$ direction. The nanorod is centrosymmetric and preserves $\mathcal{PT}$. One observes that the in-gap hinge bands connect the projections of RNLs. The hinge character is further confirmed by inspecting the real-space wave-function distribution of these in-gap modes at $k_z=0$, as shown in Fig.~\ref{fig:tb}(f). The results show a high degree of localization at two $\mathcal{PT}$-related hinges of the nanorod, which provides definitive evidence of RNLs in our TB model.

\begin{table}[t]
\caption{Parity eigenvalues for IGN at the eight $\mathcal{P}$-invariant points. $N_-$ denotes the number of occupied bands with negative parity eigenvalues. The $k_z=0$ and $k_z=\pi$ planes have $\nu_R=1$ and $\nu_R=0$, respectively. \label{table:materials}}
\begin{ruledtabular}
\begin{tabular}{l cccc c cccc}
& \multicolumn{4}{c}{$k_z=0$} & & \multicolumn{4}{c}{$k_z=\pi$} \\[1pt]
\cline{2-5} \cline{7-10}
\rule{0pt}{10pt}& $\Gamma$ & $X$ & $M$ & $Y$ & & $Z$ & $R$ & $S$ & $T$ \\[1pt]
\hline
\rule{0pt}{9pt}$N_-$ & 4 & 7 & 6 & 7 & & 6 & 6 & 6 & 6 \\[1pt]
$\nu_R$ & \multicolumn{4}{c}{1} & & \multicolumn{4}{c}{0} \\
\end{tabular}
\end{ruledtabular}
\end{table}

\emph{\color{blue}Materials realization. --- }Based on our construction scheme combined with first-principles calculations, we propose a concrete material realization --- the interpenetrated graphene network (IGN). The IGN is a three-dimensional carbon allotrope first predicted by Chen \textit{et al.} in 2015~\cite{Chen2015Nanostructured-NL}. It has been shown to host a pair of Weyl-like nodal lines protected by the quantized Berry phase. The crystal structure is shown in Fig.~\ref{fig:ign}(a). The unit cell contains six carbon atoms, forming two obtuse triangles related by inversion symmetry. The carbon atoms can be categorized into two groups based on their coordination: two fourfold-coordinated atoms with $sp^3$ hybridization and four threefold-coordinated atoms with $sp^2$ hybridization.

The structure belongs to the orthorhombic crystal system, with space group No. 63 ($Cmcm$). Both the inversion $\mathcal{P}$ and the time reversal $\mathcal{T}$ are present. More importantly, the IGN possesses a twofold screw axis $S_{2z}$ along the $z$ direction, which allows the existence of a nodal surface on the $k_z = \pi$ plane. Meanwhile, the system is essentially spinless, since the spin-orbit coupling in carbon allotropes can be neglected. These fulfill the conditions required for an RNL phase in nodal surface semimetals.

The bulk electronic band structure of IGN is shown in Fig.~\ref{fig:ign}(b). One observes that the bands form twofold-degenerate pairs along the $Z$-$R$-$S$-$Z$ path which is located on the $k_z=\pi$ plane. A careful scan for $k_z=\pi$ reveals that the two lower bands and the two upper bands are separately degenerate, forming a nodal surface. This is a direct consequence of the $S_{2z}\mathcal{T}$ symmetry. In addition, a band crossing is observed along the $\Gamma$-$Z$ path, which belongs to a pair of nodal lines traversing the BZ and located symmetrically on the two sides of the $k_z=\pi$ plane, as shown in Fig.~\ref{fig:ign}(c). The nodal lines are formed by the two middle bands, while the nodal surface is formed by the two lower (upper) bands. This reproduces the linking configuration proposed in Fig.~\ref{fig:idea}(b).

To confirm the topological character, we evaluate the real Chern number $\nu_R$ for the $k_z=0$ and $k_z=\pi$ planes using both the Wilson-loop method and the parity approach~\cite{Ahn2018Band-PRL,Chen2021Graphyne-PRB,Yue2024Stability-PRB}. The Wilson-loop spectrum on the $k_z=0$ plane is presented in Fig.~\ref{fig:ign}(e). It exhibits three crossings through $\theta=\pi$, yielding $\nu_R=1$. In contrast, the Wilson-loop spectrum on the $k_z=\pi$ plane [see Fig.~S2(c,f) in the Supplementary Material~\cite{SupplementaryMaterial}] is topologically trivial with $\nu_R=0$. The parity approach gives the same results, as deduced from the calculated values of $N_-(\Gamma_i)$ at all TRIM points listed in Table~\ref{table:materials}. Since the nodal line is the only nodal degeneracy formed by the occupied and unoccupied bands between $k_z=0$ and $k_z=\pi$ planes, the change of $\nu_R$ indicates that the nodal line carries a nontrivial real Chern number. Thus, this confirms that the nodal lines in IGN are indeed RNLs.

\begin{figure}[t]
\centering
\setlength{\abovecaptionskip}{6pt}
\includegraphics[width=\columnwidth]{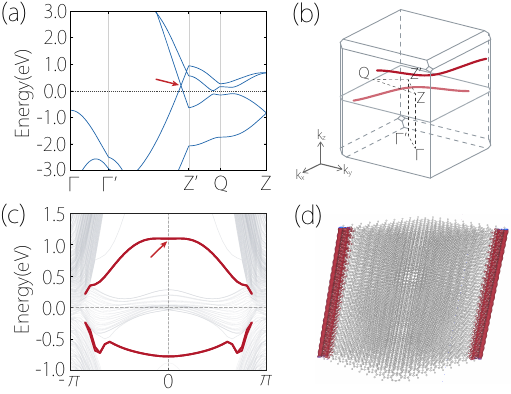}
\caption{\label{fig:screwbreaking}(a) The bulk band structure of IGN under a $10^\circ$ shear strain along the [110] direction. (b) The distorted BZ. The red lines denote the RNLs. The nodal surface is violated. (c) Nanorod spectrum ($10\times10$-cell cross section). The hinge states are marked in red. (d) Spatial distribution of the hinge mode at $k_z = 0$ [marked by the red arrow in (c)], which is localized at two $\mathcal{PT}$-related hinges.}
\end{figure}

In addition, the RNLs in IGN also feature the quantized Berry phase, and thus carry dual $\mathbb{Z}_2$ topological charges $(w,\nu_R)=(1,1)$. The nontrivial topology in the bulk manifests as topological states at both the surface and the hinge. The surface spectral function on the (010) surface is shown in Fig.~\ref{fig:ign}(d). One observes drumhead-type surface states connecting to the projected RNLs, consistent with the nontrivial 1D topological charge $w=1$. As for the topological hinge states, we construct a nanorod geometry by expanding the unit cell by $10\times10$ along the $a$ and $b$ directions with open boundaries, while keeping periodicity along the $z$ direction. The nanorod preserves $\mathcal{PT}$ symmetry. We expect that the hinge states would exist between the projections of the bulk RNLs around $k_z = 0$, since the 2D $k_z = 0$ slice maintains a nontrivial $\nu_R=1$. The energy spectrum of the nanorod is plotted in Fig.~\ref{fig:ign}(f), which indeed confirms our expectation. In Fig.~\ref{fig:ign}(g), we show the spatial distribution of charge density at $k_z=0$ for these hinge states. The results show a high degree of localization at two $\mathcal{PT}$-related hinges of the nanorod, which provides definitive evidence of RNLs in IGN.

\emph{\color{blue}Discussion. --- }
In this work, we propose a new type of RNLs in nodal surface semimetals, and predict its realization in the carbon allotrope IGN. In our construction, an $S_{2z}\mathcal{T}$-enforced nodal surface serves as the linking partner which obstructs the annihilation of a pair of RNLs. The real topology of RNLs is protected by the $\mathcal{PT}$ symmetry and is characterized by dual $\mathbb{Z}_2$ topological charges with the $\pi$ Berry phase and the nontrivial real Chern number. If the pair of RNLs meets at the nodal surface, it would form a fourfold degenerate contact instead of annihilating (see Fig.~S5 in the Supplementary Material~\cite{SupplementaryMaterial} for the evolution of RNLs in IGN by changing the lattice angle $\gamma$ that preserves $S_{2z}\mathcal{T}$). Thus, the linking configuration is completely different from the previously reported RNLs which are linked by another nodal line.

Similar to the conventional RNL, the nodal-surface-linked RNLs are also robust against symmetry-preserving perturbations. Unless the defects/impurities are strong enough to either break the $\mathcal{PT}$ symmetry or annihilate the RNLs in pairs, the state should remain stable, even if the nodal surface is removed. In Fig.~\ref{fig:screwbreaking}, we show that the RNLs in IGN are robust against a $10^\circ$ shear strain distortion that breaks the $S_{2z}$ symmetry, which removes the nodal surface but preserves $\mathcal{PT}$. The distorted system retains $\nu_R=1$ on the $k_z=0$ plane and supports hinge states. The distorted structure and its Wilson-loop spectrum are shown in Fig.~S3 of the Supplementary Material~\cite{SupplementaryMaterial}. It follows that the RNLs can still exist and retain their topological hinge Fermi arcs.

Besides the electronic bands of the materials with negligible spin-orbit coupling, our findings can also apply to many spinless systems, including phononic systems~\cite{Zhu2022Phononic-PRB,Yang2023Cladded-PRB}. Moreover, the artificial systems such as acoustic and photonic crystals may be promising platforms for realizing the RNL linked by nodal surfaces, due to the high flexibility in tuning hopping parameters~\cite{Xiao2017Topologically-arXiv,Kim2019Topologically-PRB,Xue2023Stiefel-NC,Xiang2024Demonstration-PRL,Ma2024Observation-PRL}. The proposed RNLs in nodal surface semimetals may also stimulate further studies on the interplay between real topology and other topological phases, such as the Floquet topology~\cite{Narayan2016Tunable-PRB,Ezawa2017Photoinduced-PRB,Yan2017Floquet-PRB,Du2022Weyl-PRB,Liu2025Floquet-PRB}.

\emph{\color{blue}Acknowledgements. --- }
This work is supported by the Guangdong Basic and Applied Basic Research Foundation (Grant No. 2022A1515110094), the Excellent Young Scientists Fund Program (Overseas) of China, and the program of Outstanding Young and Middle-aged Scholars of Shandong University.

% PRB requires works cited only in the Supplemental Material in the main reference list.
\nocite{Zhang2022MagneticTB-CPC,Kresse1993Ab-PRB,Kresse1996Efficient-PRB,Bloechl1994Projector-PRB,Perdew1996Generalized-PRL,Marzari1997Maximally-PRB,Souza2001Maximally-PRB,Mostofi2014An-CPC,Sancho1984Quick-JPF,Sancho1985Highly-JPF,Wu2018WannierTools-CPC}

% \bibliography{ref.bib}

\begin{thebibliography}{72}%
\makeatletter
\providecommand \@ifxundefined [1]{%
 \@ifx{#1\undefined}
}%
\providecommand \@ifnum [1]{%
 \ifnum #1\expandafter \@firstoftwo
 \else \expandafter \@secondoftwo
 \fi
}%
\providecommand \@ifx [1]{%
 \ifx #1\expandafter \@firstoftwo
 \else \expandafter \@secondoftwo
 \fi
}%
\providecommand \natexlab [1]{#1}%
\providecommand \enquote  [1]{``#1''}%
\providecommand \bibnamefont  [1]{#1}%
\providecommand \bibfnamefont [1]{#1}%
\providecommand \citenamefont [1]{#1}%
\providecommand \href@noop [0]{\@secondoftwo}%
\providecommand \href [0]{\begingroup \@sanitize@url \@href}%
\providecommand \@href[1]{\@@startlink{#1}\@@href}%
\providecommand \@@href[1]{\endgroup#1\@@endlink}%
\providecommand \@sanitize@url [0]{\catcode `\\12\catcode `\$12\catcode `\&12\catcode `\#12\catcode `\^12\catcode `\_12\catcode `\%12\relax}%
\providecommand \@@startlink[1]{}%
\providecommand \@@endlink[0]{}%
\providecommand \url  [0]{\begingroup\@sanitize@url \@url }%
\providecommand \@url [1]{\endgroup\@href {#1}{\urlprefix }}%
\providecommand \urlprefix  [0]{URL }%
\providecommand \Eprint [0]{\href }%
\providecommand \doibase [0]{https://doi.org/}%
\providecommand \selectlanguage [0]{\@gobble}%
\providecommand \bibinfo  [0]{\@secondoftwo}%
\providecommand \bibfield  [0]{\@secondoftwo}%
\providecommand \translation [1]{[#1]}%
\providecommand \BibitemOpen [0]{}%
\providecommand \bibitemStop [0]{}%
\providecommand \bibitemNoStop [0]{.\EOS\space}%
\providecommand \EOS [0]{\spacefactor3000\relax}%
\providecommand \BibitemShut  [1]{\csname bibitem#1\endcsname}%
\let\auto@bib@innerbib\@empty
%</preamble>
\bibitem [{\citenamefont {Bansil}\ \emph {et~al.}(2016)\citenamefont {Bansil}, \citenamefont {Lin},\ and\ \citenamefont {Das}}]{Bansil2016Colloquium-RMP}%
  \BibitemOpen
  \bibfield  {author} {\bibinfo {author} {\bibfnamefont {A.}~\bibnamefont {Bansil}}, \bibinfo {author} {\bibfnamefont {H.}~\bibnamefont {Lin}},\ and\ \bibinfo {author} {\bibfnamefont {T.}~\bibnamefont {Das}},\ }\bibfield  {title} {\bibinfo {title} {Colloquium: Topological band theory},\ }\href@noop {} {\bibfield  {journal} {\bibinfo  {journal} {Rev. Mod. Phys.}\ }\textbf {\bibinfo {volume} {88}},\ \bibinfo {pages} {021004} (\bibinfo {year} {2016})}\BibitemShut {NoStop}%
\bibitem [{\citenamefont {Burkov}(2016)}]{Burkov2016Topological-NM}%
  \BibitemOpen
  \bibfield  {author} {\bibinfo {author} {\bibfnamefont {A.~A.}\ \bibnamefont {Burkov}},\ }\bibfield  {title} {\bibinfo {title} {Topological semimetals},\ }\href@noop {} {\bibfield  {journal} {\bibinfo  {journal} {Nat. Mater.}\ }\textbf {\bibinfo {volume} {15}},\ \bibinfo {pages} {1145} (\bibinfo {year} {2016})}\BibitemShut {NoStop}%
\bibitem [{\citenamefont {Yang}(2016)}]{Yang2016Dirac-SPIN}%
  \BibitemOpen
  \bibfield  {author} {\bibinfo {author} {\bibfnamefont {S.~A.}\ \bibnamefont {Yang}},\ }\bibfield  {title} {\bibinfo {title} {Dirac and {Weyl} materials: fundamental aspects and some spintronics applications},\ }\href@noop {} {\bibfield  {journal} {\bibinfo  {journal} {SPIN}\ }\textbf {\bibinfo {volume} {6}},\ \bibinfo {pages} {1640003} (\bibinfo {year} {2016})}\BibitemShut {NoStop}%
\bibitem [{\citenamefont {Armitage}\ \emph {et~al.}(2018)\citenamefont {Armitage}, \citenamefont {Mele},\ and\ \citenamefont {Vishwanath}}]{Armitage2018Weyl-RMP}%
  \BibitemOpen
  \bibfield  {author} {\bibinfo {author} {\bibfnamefont {N.~P.}\ \bibnamefont {Armitage}}, \bibinfo {author} {\bibfnamefont {E.~J.}\ \bibnamefont {Mele}},\ and\ \bibinfo {author} {\bibfnamefont {A.}~\bibnamefont {Vishwanath}},\ }\bibfield  {title} {\bibinfo {title} {Weyl and {Dirac} semimetals in three-dimensional solids},\ }\href@noop {} {\bibfield  {journal} {\bibinfo  {journal} {Rev. Mod. Phys.}\ }\textbf {\bibinfo {volume} {90}},\ \bibinfo {pages} {015001} (\bibinfo {year} {2018})}\BibitemShut {NoStop}%
\bibitem [{\citenamefont {Yu}\ \emph {et~al.}(2022)\citenamefont {Yu}, \citenamefont {Zhang}, \citenamefont {Liu}, \citenamefont {Wu}, \citenamefont {Li}, \citenamefont {Zhang}, \citenamefont {Yang},\ and\ \citenamefont {Yao}}]{Yu2022Encyclopedia-SB}%
  \BibitemOpen
  \bibfield  {author} {\bibinfo {author} {\bibfnamefont {Z.-M.}\ \bibnamefont {Yu}}, \bibinfo {author} {\bibfnamefont {Z.}~\bibnamefont {Zhang}}, \bibinfo {author} {\bibfnamefont {G.-B.}\ \bibnamefont {Liu}}, \bibinfo {author} {\bibfnamefont {W.}~\bibnamefont {Wu}}, \bibinfo {author} {\bibfnamefont {X.-P.}\ \bibnamefont {Li}}, \bibinfo {author} {\bibfnamefont {R.-W.}\ \bibnamefont {Zhang}}, \bibinfo {author} {\bibfnamefont {S.~A.}\ \bibnamefont {Yang}},\ and\ \bibinfo {author} {\bibfnamefont {Y.}~\bibnamefont {Yao}},\ }\bibfield  {title} {\bibinfo {title} {Encyclopedia of emergent particles in three-dimensional crystals},\ }\href@noop {} {\bibfield  {journal} {\bibinfo  {journal} {Sci. Bull.}\ }\textbf {\bibinfo {volume} {67}},\ \bibinfo {pages} {375} (\bibinfo {year} {2022})}\BibitemShut {NoStop}%
\bibitem [{\citenamefont {Wan}\ \emph {et~al.}(2011)\citenamefont {Wan}, \citenamefont {Turner}, \citenamefont {Vishwanath},\ and\ \citenamefont {Savrasov}}]{Wan2011Topological-PRB}%
  \BibitemOpen
  \bibfield  {author} {\bibinfo {author} {\bibfnamefont {X.}~\bibnamefont {Wan}}, \bibinfo {author} {\bibfnamefont {A.~M.}\ \bibnamefont {Turner}}, \bibinfo {author} {\bibfnamefont {A.}~\bibnamefont {Vishwanath}},\ and\ \bibinfo {author} {\bibfnamefont {S.~Y.}\ \bibnamefont {Savrasov}},\ }\bibfield  {title} {\bibinfo {title} {Topological semimetal and {Fermi}-arc surface states in the electronic structure of pyrochlore iridates},\ }\href@noop {} {\bibfield  {journal} {\bibinfo  {journal} {Phys. Rev. B}\ }\textbf {\bibinfo {volume} {83}},\ \bibinfo {pages} {205101} (\bibinfo {year} {2011})}\BibitemShut {NoStop}%
\bibitem [{\citenamefont {Xu}\ \emph {et~al.}(2011)\citenamefont {Xu}, \citenamefont {Weng}, \citenamefont {Wang}, \citenamefont {Dai},\ and\ \citenamefont {Fang}}]{Xu2011Chern-PRL}%
  \BibitemOpen
  \bibfield  {author} {\bibinfo {author} {\bibfnamefont {G.}~\bibnamefont {Xu}}, \bibinfo {author} {\bibfnamefont {H.}~\bibnamefont {Weng}}, \bibinfo {author} {\bibfnamefont {Z.}~\bibnamefont {Wang}}, \bibinfo {author} {\bibfnamefont {X.}~\bibnamefont {Dai}},\ and\ \bibinfo {author} {\bibfnamefont {Z.}~\bibnamefont {Fang}},\ }\bibfield  {title} {\bibinfo {title} {Chern semimetal and the quantized anomalous {Hall} effect in {HgCr$_2$Se$_4$}},\ }\href@noop {} {\bibfield  {journal} {\bibinfo  {journal} {Phys. Rev. Lett.}\ }\textbf {\bibinfo {volume} {107}},\ \bibinfo {pages} {186806} (\bibinfo {year} {2011})}\BibitemShut {NoStop}%
\bibitem [{\citenamefont {Fang}\ \emph {et~al.}(2012)\citenamefont {Fang}, \citenamefont {Gilbert}, \citenamefont {Dai},\ and\ \citenamefont {Bernevig}}]{Fang2012Multi-PRL}%
  \BibitemOpen
  \bibfield  {author} {\bibinfo {author} {\bibfnamefont {C.}~\bibnamefont {Fang}}, \bibinfo {author} {\bibfnamefont {M.~J.}\ \bibnamefont {Gilbert}}, \bibinfo {author} {\bibfnamefont {X.}~\bibnamefont {Dai}},\ and\ \bibinfo {author} {\bibfnamefont {B.~A.}\ \bibnamefont {Bernevig}},\ }\bibfield  {title} {\bibinfo {title} {{Multi-Weyl} topological semimetals stabilized by point group symmetry},\ }\href@noop {} {\bibfield  {journal} {\bibinfo  {journal} {Phys. Rev. Lett.}\ }\textbf {\bibinfo {volume} {108}},\ \bibinfo {pages} {266802} (\bibinfo {year} {2012})}\BibitemShut {NoStop}%
\bibitem [{\citenamefont {Young}\ \emph {et~al.}(2012)\citenamefont {Young}, \citenamefont {Zaheer}, \citenamefont {Teo}, \citenamefont {Kane}, \citenamefont {Mele},\ and\ \citenamefont {Rappe}}]{Young2012Dirac-PRL}%
  \BibitemOpen
  \bibfield  {author} {\bibinfo {author} {\bibfnamefont {S.~M.}\ \bibnamefont {Young}}, \bibinfo {author} {\bibfnamefont {S.}~\bibnamefont {Zaheer}}, \bibinfo {author} {\bibfnamefont {J.~C.~Y.}\ \bibnamefont {Teo}}, \bibinfo {author} {\bibfnamefont {C.~L.}\ \bibnamefont {Kane}}, \bibinfo {author} {\bibfnamefont {E.~J.}\ \bibnamefont {Mele}},\ and\ \bibinfo {author} {\bibfnamefont {A.~M.}\ \bibnamefont {Rappe}},\ }\bibfield  {title} {\bibinfo {title} {Dirac semimetal in three dimensions},\ }\href@noop {} {\bibfield  {journal} {\bibinfo  {journal} {Phys. Rev. Lett.}\ }\textbf {\bibinfo {volume} {108}},\ \bibinfo {pages} {140405} (\bibinfo {year} {2012})}\BibitemShut {NoStop}%
\bibitem [{\citenamefont {Wang}\ \emph {et~al.}(2012)\citenamefont {Wang}, \citenamefont {Sun}, \citenamefont {Chen}, \citenamefont {Franchini}, \citenamefont {Xu}, \citenamefont {Weng}, \citenamefont {Dai},\ and\ \citenamefont {Fang}}]{Wang2012Dirac-PRB}%
  \BibitemOpen
  \bibfield  {author} {\bibinfo {author} {\bibfnamefont {Z.}~\bibnamefont {Wang}}, \bibinfo {author} {\bibfnamefont {Y.}~\bibnamefont {Sun}}, \bibinfo {author} {\bibfnamefont {X.-Q.}\ \bibnamefont {Chen}}, \bibinfo {author} {\bibfnamefont {C.}~\bibnamefont {Franchini}}, \bibinfo {author} {\bibfnamefont {G.}~\bibnamefont {Xu}}, \bibinfo {author} {\bibfnamefont {H.}~\bibnamefont {Weng}}, \bibinfo {author} {\bibfnamefont {X.}~\bibnamefont {Dai}},\ and\ \bibinfo {author} {\bibfnamefont {Z.}~\bibnamefont {Fang}},\ }\bibfield  {title} {\bibinfo {title} {Dirac semimetal and topological phase transitions in {$A_3$Bi} ({A} = {Na}, {K}, {Rb})},\ }\href@noop {} {\bibfield  {journal} {\bibinfo  {journal} {Phys. Rev. B}\ }\textbf {\bibinfo {volume} {85}},\ \bibinfo {pages} {195320} (\bibinfo {year} {2012})}\BibitemShut {NoStop}%
\bibitem [{\citenamefont {Wang}\ \emph {et~al.}(2013)\citenamefont {Wang}, \citenamefont {Weng}, \citenamefont {Wu}, \citenamefont {Dai},\ and\ \citenamefont {Fang}}]{Wang2013Three-PRB}%
  \BibitemOpen
  \bibfield  {author} {\bibinfo {author} {\bibfnamefont {Z.}~\bibnamefont {Wang}}, \bibinfo {author} {\bibfnamefont {H.}~\bibnamefont {Weng}}, \bibinfo {author} {\bibfnamefont {Q.}~\bibnamefont {Wu}}, \bibinfo {author} {\bibfnamefont {X.}~\bibnamefont {Dai}},\ and\ \bibinfo {author} {\bibfnamefont {Z.}~\bibnamefont {Fang}},\ }\bibfield  {title} {\bibinfo {title} {Three-dimensional {Dirac} semimetal and quantum transport in {Cd$_3$As$_2$}},\ }\href@noop {} {\bibfield  {journal} {\bibinfo  {journal} {Phys. Rev. B}\ }\textbf {\bibinfo {volume} {88}},\ \bibinfo {pages} {125427} (\bibinfo {year} {2013})}\BibitemShut {NoStop}%
\bibitem [{\citenamefont {Weng}\ \emph {et~al.}(2015)\citenamefont {Weng}, \citenamefont {Liang}, \citenamefont {Xu}, \citenamefont {Yu}, \citenamefont {Fang}, \citenamefont {Dai},\ and\ \citenamefont {Kawazoe}}]{Weng2015Topological-PRB}%
  \BibitemOpen
  \bibfield  {author} {\bibinfo {author} {\bibfnamefont {H.}~\bibnamefont {Weng}}, \bibinfo {author} {\bibfnamefont {Y.}~\bibnamefont {Liang}}, \bibinfo {author} {\bibfnamefont {Q.}~\bibnamefont {Xu}}, \bibinfo {author} {\bibfnamefont {R.}~\bibnamefont {Yu}}, \bibinfo {author} {\bibfnamefont {Z.}~\bibnamefont {Fang}}, \bibinfo {author} {\bibfnamefont {X.}~\bibnamefont {Dai}},\ and\ \bibinfo {author} {\bibfnamefont {Y.}~\bibnamefont {Kawazoe}},\ }\bibfield  {title} {\bibinfo {title} {Topological node-line semimetal in three-dimensional graphene networks},\ }\href@noop {} {\bibfield  {journal} {\bibinfo  {journal} {Phys. Rev. B}\ }\textbf {\bibinfo {volume} {92}},\ \bibinfo {pages} {045108} (\bibinfo {year} {2015})}\BibitemShut {NoStop}%
\bibitem [{\citenamefont {Chen}\ \emph {et~al.}(2015)\citenamefont {Chen}, \citenamefont {Xie}, \citenamefont {Yang}, \citenamefont {Pan}, \citenamefont {Zhang}, \citenamefont {Cohen},\ and\ \citenamefont {Zhang}}]{Chen2015Nanostructured-NL}%
  \BibitemOpen
  \bibfield  {author} {\bibinfo {author} {\bibfnamefont {Y.}~\bibnamefont {Chen}}, \bibinfo {author} {\bibfnamefont {Y.}~\bibnamefont {Xie}}, \bibinfo {author} {\bibfnamefont {S.~A.}\ \bibnamefont {Yang}}, \bibinfo {author} {\bibfnamefont {H.}~\bibnamefont {Pan}}, \bibinfo {author} {\bibfnamefont {F.}~\bibnamefont {Zhang}}, \bibinfo {author} {\bibfnamefont {M.~L.}\ \bibnamefont {Cohen}},\ and\ \bibinfo {author} {\bibfnamefont {S.}~\bibnamefont {Zhang}},\ }\bibfield  {title} {\bibinfo {title} {Nanostructured carbon allotropes with {Weyl}-like loops and points},\ }\href@noop {} {\bibfield  {journal} {\bibinfo  {journal} {Nano Lett.}\ }\textbf {\bibinfo {volume} {15}},\ \bibinfo {pages} {6974} (\bibinfo {year} {2015})}\BibitemShut {NoStop}%
\bibitem [{\citenamefont {Kim}\ \emph {et~al.}(2015)\citenamefont {Kim}, \citenamefont {Wieder}, \citenamefont {Kane},\ and\ \citenamefont {Rappe}}]{Kim2015Dirac-PRL}%
  \BibitemOpen
  \bibfield  {author} {\bibinfo {author} {\bibfnamefont {Y.}~\bibnamefont {Kim}}, \bibinfo {author} {\bibfnamefont {B.~J.}\ \bibnamefont {Wieder}}, \bibinfo {author} {\bibfnamefont {C.~L.}\ \bibnamefont {Kane}},\ and\ \bibinfo {author} {\bibfnamefont {A.~M.}\ \bibnamefont {Rappe}},\ }\bibfield  {title} {\bibinfo {title} {Dirac line nodes in inversion-symmetric crystals},\ }\href@noop {} {\bibfield  {journal} {\bibinfo  {journal} {Phys. Rev. Lett.}\ }\textbf {\bibinfo {volume} {115}},\ \bibinfo {pages} {036806} (\bibinfo {year} {2015})}\BibitemShut {NoStop}%
\bibitem [{\citenamefont {Yu}\ \emph {et~al.}(2015)\citenamefont {Yu}, \citenamefont {Weng}, \citenamefont {Fang}, \citenamefont {Dai},\ and\ \citenamefont {Hu}}]{Yu2015Topological-PRL}%
  \BibitemOpen
  \bibfield  {author} {\bibinfo {author} {\bibfnamefont {R.}~\bibnamefont {Yu}}, \bibinfo {author} {\bibfnamefont {H.}~\bibnamefont {Weng}}, \bibinfo {author} {\bibfnamefont {Z.}~\bibnamefont {Fang}}, \bibinfo {author} {\bibfnamefont {X.}~\bibnamefont {Dai}},\ and\ \bibinfo {author} {\bibfnamefont {X.}~\bibnamefont {Hu}},\ }\bibfield  {title} {\bibinfo {title} {Topological node-line semimetal and {Dirac} semimetal state in antiperovskite {${\mathrm{Cu}}_{3}\mathrm{PdN}$}},\ }\href@noop {} {\bibfield  {journal} {\bibinfo  {journal} {Phys. Rev. Lett.}\ }\textbf {\bibinfo {volume} {115}},\ \bibinfo {pages} {036807} (\bibinfo {year} {2015})}\BibitemShut {NoStop}%
\bibitem [{\citenamefont {Fang}\ \emph {et~al.}(2016)\citenamefont {Fang}, \citenamefont {Weng}, \citenamefont {Dai},\ and\ \citenamefont {Fang}}]{Fang2016Topological-CPB}%
  \BibitemOpen
  \bibfield  {author} {\bibinfo {author} {\bibfnamefont {C.}~\bibnamefont {Fang}}, \bibinfo {author} {\bibfnamefont {H.}~\bibnamefont {Weng}}, \bibinfo {author} {\bibfnamefont {X.}~\bibnamefont {Dai}},\ and\ \bibinfo {author} {\bibfnamefont {Z.}~\bibnamefont {Fang}},\ }\bibfield  {title} {\bibinfo {title} {Topological nodal line semimetals},\ }\href@noop {} {\bibfield  {journal} {\bibinfo  {journal} {Chin. Phys. B}\ }\textbf {\bibinfo {volume} {25}},\ \bibinfo {pages} {117106} (\bibinfo {year} {2016})}\BibitemShut {NoStop}%
\bibitem [{\citenamefont {Yu}\ \emph {et~al.}(2019)\citenamefont {Yu}, \citenamefont {Wu}, \citenamefont {Sheng}, \citenamefont {Zhao},\ and\ \citenamefont {Yang}}]{Yu2019Quadratic-PRB}%
  \BibitemOpen
  \bibfield  {author} {\bibinfo {author} {\bibfnamefont {Z.-M.}\ \bibnamefont {Yu}}, \bibinfo {author} {\bibfnamefont {W.}~\bibnamefont {Wu}}, \bibinfo {author} {\bibfnamefont {X.-L.}\ \bibnamefont {Sheng}}, \bibinfo {author} {\bibfnamefont {Y.~X.}\ \bibnamefont {Zhao}},\ and\ \bibinfo {author} {\bibfnamefont {S.~A.}\ \bibnamefont {Yang}},\ }\bibfield  {title} {\bibinfo {title} {Quadratic and cubic nodal lines stabilized by crystalline symmetry},\ }\href@noop {} {\bibfield  {journal} {\bibinfo  {journal} {Phys. Rev. B}\ }\textbf {\bibinfo {volume} {99}},\ \bibinfo {pages} {121106(R)} (\bibinfo {year} {2019})}\BibitemShut {NoStop}%
\bibitem [{\citenamefont {Liang}\ \emph {et~al.}(2016)\citenamefont {Liang}, \citenamefont {Zhou}, \citenamefont {Yu}, \citenamefont {Wang},\ and\ \citenamefont {Weng}}]{Liang2016Nodal-PRB}%
  \BibitemOpen
  \bibfield  {author} {\bibinfo {author} {\bibfnamefont {Q.-F.}\ \bibnamefont {Liang}}, \bibinfo {author} {\bibfnamefont {J.}~\bibnamefont {Zhou}}, \bibinfo {author} {\bibfnamefont {R.}~\bibnamefont {Yu}}, \bibinfo {author} {\bibfnamefont {Z.}~\bibnamefont {Wang}},\ and\ \bibinfo {author} {\bibfnamefont {H.}~\bibnamefont {Weng}},\ }\bibfield  {title} {\bibinfo {title} {Node-surface and node-line fermions from nonsymmorphic lattice symmetries},\ }\href@noop {} {\bibfield  {journal} {\bibinfo  {journal} {Phys. Rev. B}\ }\textbf {\bibinfo {volume} {93}},\ \bibinfo {pages} {085427} (\bibinfo {year} {2016})}\BibitemShut {NoStop}%
\bibitem [{\citenamefont {Zhong}\ \emph {et~al.}(2016)\citenamefont {Zhong}, \citenamefont {Chen}, \citenamefont {Xie}, \citenamefont {Yang}, \citenamefont {Cohen},\ and\ \citenamefont {Zhang}}]{Zhong2016Towards-N}%
  \BibitemOpen
  \bibfield  {author} {\bibinfo {author} {\bibfnamefont {C.}~\bibnamefont {Zhong}}, \bibinfo {author} {\bibfnamefont {Y.}~\bibnamefont {Chen}}, \bibinfo {author} {\bibfnamefont {Y.}~\bibnamefont {Xie}}, \bibinfo {author} {\bibfnamefont {S.~A.}\ \bibnamefont {Yang}}, \bibinfo {author} {\bibfnamefont {M.~L.}\ \bibnamefont {Cohen}},\ and\ \bibinfo {author} {\bibfnamefont {S.~B.}\ \bibnamefont {Zhang}},\ }\bibfield  {title} {\bibinfo {title} {Towards three-dimensional {Weyl}-surface semimetals in graphene networks},\ }\href@noop {} {\bibfield  {journal} {\bibinfo  {journal} {Nanoscale}\ }\textbf {\bibinfo {volume} {8}},\ \bibinfo {pages} {7232} (\bibinfo {year} {2016})}\BibitemShut {NoStop}%
\bibitem [{\citenamefont {Wu}\ \emph {et~al.}(2018{\natexlab{a}})\citenamefont {Wu}, \citenamefont {Liu}, \citenamefont {Li}, \citenamefont {Zhong}, \citenamefont {Yu}, \citenamefont {Sheng}, \citenamefont {Zhao},\ and\ \citenamefont {Yang}}]{Wu2018Nodal-PRB}%
  \BibitemOpen
  \bibfield  {author} {\bibinfo {author} {\bibfnamefont {W.}~\bibnamefont {Wu}}, \bibinfo {author} {\bibfnamefont {Y.}~\bibnamefont {Liu}}, \bibinfo {author} {\bibfnamefont {S.}~\bibnamefont {Li}}, \bibinfo {author} {\bibfnamefont {C.}~\bibnamefont {Zhong}}, \bibinfo {author} {\bibfnamefont {Z.-M.}\ \bibnamefont {Yu}}, \bibinfo {author} {\bibfnamefont {X.-L.}\ \bibnamefont {Sheng}}, \bibinfo {author} {\bibfnamefont {Y.~X.}\ \bibnamefont {Zhao}},\ and\ \bibinfo {author} {\bibfnamefont {S.~A.}\ \bibnamefont {Yang}},\ }\bibfield  {title} {\bibinfo {title} {Nodal surface semimetals: Theory and material realization},\ }\href@noop {} {\bibfield  {journal} {\bibinfo  {journal} {Phys. Rev. B}\ }\textbf {\bibinfo {volume} {97}},\ \bibinfo {pages} {115125} (\bibinfo {year} {2018}{\natexlab{a}})}\BibitemShut {NoStop}%
\bibitem [{\citenamefont {Bzdušek}\ \emph {et~al.}(2016)\citenamefont {Bzdušek}, \citenamefont {Wu}, \citenamefont {Rüegg}, \citenamefont {Sigrist},\ and\ \citenamefont {Soluyanov}}]{Bzdusek2016Nodal-N}%
  \BibitemOpen
  \bibfield  {author} {\bibinfo {author} {\bibfnamefont {T.}~\bibnamefont {Bzdušek}}, \bibinfo {author} {\bibfnamefont {Q.}~\bibnamefont {Wu}}, \bibinfo {author} {\bibfnamefont {A.}~\bibnamefont {Rüegg}}, \bibinfo {author} {\bibfnamefont {M.}~\bibnamefont {Sigrist}},\ and\ \bibinfo {author} {\bibfnamefont {A.~A.}\ \bibnamefont {Soluyanov}},\ }\bibfield  {title} {\bibinfo {title} {Nodal-chain metals},\ }\href@noop {} {\bibfield  {journal} {\bibinfo  {journal} {Nature}\ }\textbf {\bibinfo {volume} {538}},\ \bibinfo {pages} {75} (\bibinfo {year} {2016})}\BibitemShut {NoStop}%
\bibitem [{\citenamefont {Yu}\ \emph {et~al.}(2017)\citenamefont {Yu}, \citenamefont {Wu}, \citenamefont {Fang},\ and\ \citenamefont {Weng}}]{Yu2017From-PRL}%
  \BibitemOpen
  \bibfield  {author} {\bibinfo {author} {\bibfnamefont {R.}~\bibnamefont {Yu}}, \bibinfo {author} {\bibfnamefont {Q.}~\bibnamefont {Wu}}, \bibinfo {author} {\bibfnamefont {Z.}~\bibnamefont {Fang}},\ and\ \bibinfo {author} {\bibfnamefont {H.}~\bibnamefont {Weng}},\ }\bibfield  {title} {\bibinfo {title} {From nodal chain semimetal to {Weyl} semimetal in {HfC}},\ }\href@noop {} {\bibfield  {journal} {\bibinfo  {journal} {Phys. Rev. Lett.}\ }\textbf {\bibinfo {volume} {119}},\ \bibinfo {pages} {036401} (\bibinfo {year} {2017})}\BibitemShut {NoStop}%
\bibitem [{\citenamefont {Wang}\ \emph {et~al.}(2017)\citenamefont {Wang}, \citenamefont {Liu}, \citenamefont {Yu}, \citenamefont {Sheng},\ and\ \citenamefont {Yang}}]{Wang2017Hourglass-NC}%
  \BibitemOpen
  \bibfield  {author} {\bibinfo {author} {\bibfnamefont {S.-S.}\ \bibnamefont {Wang}}, \bibinfo {author} {\bibfnamefont {Y.}~\bibnamefont {Liu}}, \bibinfo {author} {\bibfnamefont {Z.-M.}\ \bibnamefont {Yu}}, \bibinfo {author} {\bibfnamefont {X.-L.}\ \bibnamefont {Sheng}},\ and\ \bibinfo {author} {\bibfnamefont {S.~A.}\ \bibnamefont {Yang}},\ }\bibfield  {title} {\bibinfo {title} {Hourglass {Dirac} chain metal in rhenium dioxide},\ }\href@noop {} {\bibfield  {journal} {\bibinfo  {journal} {Nat. Commun.}\ }\textbf {\bibinfo {volume} {8}},\ \bibinfo {pages} {1844} (\bibinfo {year} {2017})}\BibitemShut {NoStop}%
\bibitem [{\citenamefont {Chang}\ \emph {et~al.}(2017)\citenamefont {Chang}, \citenamefont {Xu}, \citenamefont {Zhou}, \citenamefont {Huang}, \citenamefont {Singh}, \citenamefont {Wang}, \citenamefont {Belopolski}, \citenamefont {Yin}, \citenamefont {Zhang}, \citenamefont {Bansil}, \citenamefont {Lin},\ and\ \citenamefont {Hasan}}]{Chang2017Topological-PRL}%
  \BibitemOpen
  \bibfield  {author} {\bibinfo {author} {\bibfnamefont {G.}~\bibnamefont {Chang}}, \bibinfo {author} {\bibfnamefont {S.-Y.}\ \bibnamefont {Xu}}, \bibinfo {author} {\bibfnamefont {X.}~\bibnamefont {Zhou}}, \bibinfo {author} {\bibfnamefont {S.-M.}\ \bibnamefont {Huang}}, \bibinfo {author} {\bibfnamefont {B.}~\bibnamefont {Singh}}, \bibinfo {author} {\bibfnamefont {B.}~\bibnamefont {Wang}}, \bibinfo {author} {\bibfnamefont {I.}~\bibnamefont {Belopolski}}, \bibinfo {author} {\bibfnamefont {J.}~\bibnamefont {Yin}}, \bibinfo {author} {\bibfnamefont {S.}~\bibnamefont {Zhang}}, \bibinfo {author} {\bibfnamefont {A.}~\bibnamefont {Bansil}}, \bibinfo {author} {\bibfnamefont {H.}~\bibnamefont {Lin}},\ and\ \bibinfo {author} {\bibfnamefont {M.~Z.}\ \bibnamefont {Hasan}},\ }\bibfield  {title} {\bibinfo {title} {Topological {Hopf} and chain link semimetal states and their application to {Co$_2$MnGa}},\ }\href@noop {} {\bibfield  {journal} {\bibinfo  {journal} {Phys. Rev. Lett.}\ }\textbf {\bibinfo {volume} {119}},\ \bibinfo {pages} {156401} (\bibinfo {year} {2017})}\BibitemShut {NoStop}%
\bibitem [{\citenamefont {Yan}\ \emph {et~al.}(2017)\citenamefont {Yan}, \citenamefont {Bi}, \citenamefont {Shen}, \citenamefont {Lu}, \citenamefont {Zhang},\ and\ \citenamefont {Wang}}]{Yan2017Nodal-PRB}%
  \BibitemOpen
  \bibfield  {author} {\bibinfo {author} {\bibfnamefont {Z.}~\bibnamefont {Yan}}, \bibinfo {author} {\bibfnamefont {R.}~\bibnamefont {Bi}}, \bibinfo {author} {\bibfnamefont {H.}~\bibnamefont {Shen}}, \bibinfo {author} {\bibfnamefont {L.}~\bibnamefont {Lu}}, \bibinfo {author} {\bibfnamefont {S.-C.}\ \bibnamefont {Zhang}},\ and\ \bibinfo {author} {\bibfnamefont {Z.}~\bibnamefont {Wang}},\ }\bibfield  {title} {\bibinfo {title} {Nodal-link semimetals},\ }\href@noop {} {\bibfield  {journal} {\bibinfo  {journal} {Phys. Rev. B}\ }\textbf {\bibinfo {volume} {96}},\ \bibinfo {pages} {041103} (\bibinfo {year} {2017})}\BibitemShut {NoStop}%
\bibitem [{\citenamefont {Bi}\ \emph {et~al.}(2017)\citenamefont {Bi}, \citenamefont {Yan}, \citenamefont {Lu},\ and\ \citenamefont {Wang}}]{Bi2017Nodal-PRB}%
  \BibitemOpen
  \bibfield  {author} {\bibinfo {author} {\bibfnamefont {R.}~\bibnamefont {Bi}}, \bibinfo {author} {\bibfnamefont {Z.}~\bibnamefont {Yan}}, \bibinfo {author} {\bibfnamefont {L.}~\bibnamefont {Lu}},\ and\ \bibinfo {author} {\bibfnamefont {Z.}~\bibnamefont {Wang}},\ }\bibfield  {title} {\bibinfo {title} {Nodal-knot semimetals},\ }\href@noop {} {\bibfield  {journal} {\bibinfo  {journal} {Phys. Rev. B}\ }\textbf {\bibinfo {volume} {96}},\ \bibinfo {pages} {201305} (\bibinfo {year} {2017})}\BibitemShut {NoStop}%
\bibitem [{\citenamefont {Morimoto}\ and\ \citenamefont {Furusaki}(2014)}]{Morimoto2014Weyl-PRB}%
  \BibitemOpen
  \bibfield  {author} {\bibinfo {author} {\bibfnamefont {T.}~\bibnamefont {Morimoto}}\ and\ \bibinfo {author} {\bibfnamefont {A.}~\bibnamefont {Furusaki}},\ }\bibfield  {title} {\bibinfo {title} {Weyl and {Dirac} semimetals with {$\mathbb{Z}_2$} topological charge},\ }\href@noop {} {\bibfield  {journal} {\bibinfo  {journal} {Phys. Rev. B}\ }\textbf {\bibinfo {volume} {89}},\ \bibinfo {pages} {235127} (\bibinfo {year} {2014})}\BibitemShut {NoStop}%
\bibitem [{\citenamefont {Fang}\ \emph {et~al.}(2015)\citenamefont {Fang}, \citenamefont {Chen}, \citenamefont {Kee},\ and\ \citenamefont {Fu}}]{Fang2015Topological-PRB}%
  \BibitemOpen
  \bibfield  {author} {\bibinfo {author} {\bibfnamefont {C.}~\bibnamefont {Fang}}, \bibinfo {author} {\bibfnamefont {Y.}~\bibnamefont {Chen}}, \bibinfo {author} {\bibfnamefont {H.-Y.}\ \bibnamefont {Kee}},\ and\ \bibinfo {author} {\bibfnamefont {L.}~\bibnamefont {Fu}},\ }\bibfield  {title} {\bibinfo {title} {Topological nodal line semimetals with and without spin-orbital coupling},\ }\href@noop {} {\bibfield  {journal} {\bibinfo  {journal} {Phys. Rev. B}\ }\textbf {\bibinfo {volume} {92}},\ \bibinfo {pages} {081201} (\bibinfo {year} {2015})}\BibitemShut {NoStop}%
\bibitem [{\citenamefont {Zhao}\ \emph {et~al.}(2016)\citenamefont {Zhao}, \citenamefont {Schnyder},\ and\ \citenamefont {Wang}}]{Zhao2016Unified-PRL}%
  \BibitemOpen
  \bibfield  {author} {\bibinfo {author} {\bibfnamefont {Y.~X.}\ \bibnamefont {Zhao}}, \bibinfo {author} {\bibfnamefont {A.~P.}\ \bibnamefont {Schnyder}},\ and\ \bibinfo {author} {\bibfnamefont {Z.~D.}\ \bibnamefont {Wang}},\ }\bibfield  {title} {\bibinfo {title} {Unified theory of {PT} and {CP} invariant topological metals and nodal superconductors},\ }\href@noop {} {\bibfield  {journal} {\bibinfo  {journal} {Phys. Rev. Lett.}\ }\textbf {\bibinfo {volume} {116}},\ \bibinfo {pages} {156402} (\bibinfo {year} {2016})}\BibitemShut {NoStop}%
\bibitem [{\citenamefont {Zhao}\ and\ \citenamefont {Lu}(2017)}]{Zhao2017PT-PRL}%
  \BibitemOpen
  \bibfield  {author} {\bibinfo {author} {\bibfnamefont {Y.~X.}\ \bibnamefont {Zhao}}\ and\ \bibinfo {author} {\bibfnamefont {Y.}~\bibnamefont {Lu}},\ }\bibfield  {title} {\bibinfo {title} {{PT}-symmetric real {Dirac} fermions and semimetals},\ }\href@noop {} {\bibfield  {journal} {\bibinfo  {journal} {Phys. Rev. Lett.}\ }\textbf {\bibinfo {volume} {118}},\ \bibinfo {pages} {056401} (\bibinfo {year} {2017})}\BibitemShut {NoStop}%
\bibitem [{\citenamefont {Ahn}\ \emph {et~al.}(2018)\citenamefont {Ahn}, \citenamefont {Kim}, \citenamefont {Kim},\ and\ \citenamefont {Yang}}]{Ahn2018Band-PRL}%
  \BibitemOpen
  \bibfield  {author} {\bibinfo {author} {\bibfnamefont {J.}~\bibnamefont {Ahn}}, \bibinfo {author} {\bibfnamefont {D.}~\bibnamefont {Kim}}, \bibinfo {author} {\bibfnamefont {Y.}~\bibnamefont {Kim}},\ and\ \bibinfo {author} {\bibfnamefont {B.-J.}\ \bibnamefont {Yang}},\ }\bibfield  {title} {\bibinfo {title} {Band topology and linking structure of nodal line semimetals with {$\mathbb{Z}_2$} monopole charges},\ }\href@noop {} {\bibfield  {journal} {\bibinfo  {journal} {Phys. Rev. Lett.}\ }\textbf {\bibinfo {volume} {121}},\ \bibinfo {pages} {106403} (\bibinfo {year} {2018})}\BibitemShut {NoStop}%
\bibitem [{\citenamefont {Ahn}\ \emph {et~al.}(2019)\citenamefont {Ahn}, \citenamefont {Park}, \citenamefont {Kim}, \citenamefont {Kim},\ and\ \citenamefont {Yang}}]{Ahn2019Stiefel-CPB}%
  \BibitemOpen
  \bibfield  {author} {\bibinfo {author} {\bibfnamefont {J.}~\bibnamefont {Ahn}}, \bibinfo {author} {\bibfnamefont {S.}~\bibnamefont {Park}}, \bibinfo {author} {\bibfnamefont {D.}~\bibnamefont {Kim}}, \bibinfo {author} {\bibfnamefont {Y.}~\bibnamefont {Kim}},\ and\ \bibinfo {author} {\bibfnamefont {B.-J.}\ \bibnamefont {Yang}},\ }\bibfield  {title} {\bibinfo {title} {{Stiefel-Whitney} classes and topological phases in band theory},\ }\href@noop {} {\bibfield  {journal} {\bibinfo  {journal} {Chin. Phys. B}\ }\textbf {\bibinfo {volume} {28}},\ \bibinfo {pages} {117101} (\bibinfo {year} {2019})}\BibitemShut {NoStop}%
\bibitem [{\citenamefont {Wang}\ \emph {et~al.}(2020)\citenamefont {Wang}, \citenamefont {Dai}, \citenamefont {Shao}, \citenamefont {Yang},\ and\ \citenamefont {Zhao}}]{Wang2020Boundary-PRL}%
  \BibitemOpen
  \bibfield  {author} {\bibinfo {author} {\bibfnamefont {K.}~\bibnamefont {Wang}}, \bibinfo {author} {\bibfnamefont {J.-X.}\ \bibnamefont {Dai}}, \bibinfo {author} {\bibfnamefont {L.~B.}\ \bibnamefont {Shao}}, \bibinfo {author} {\bibfnamefont {S.~A.}\ \bibnamefont {Yang}},\ and\ \bibinfo {author} {\bibfnamefont {Y.~X.}\ \bibnamefont {Zhao}},\ }\bibfield  {title} {\bibinfo {title} {Boundary criticality of {PT}-invariant topology and second-order nodal-line semimetals},\ }\href@noop {} {\bibfield  {journal} {\bibinfo  {journal} {Phys. Rev. Lett.}\ }\textbf {\bibinfo {volume} {125}},\ \bibinfo {pages} {126403} (\bibinfo {year} {2020})}\BibitemShut {NoStop}%
\bibitem [{\citenamefont {Chen}\ \emph {et~al.}(2021)\citenamefont {Chen}, \citenamefont {Wu}, \citenamefont {Yu}, \citenamefont {Chen}, \citenamefont {Zhao}, \citenamefont {Sheng},\ and\ \citenamefont {Yang}}]{Chen2021Graphyne-PRB}%
  \BibitemOpen
  \bibfield  {author} {\bibinfo {author} {\bibfnamefont {C.}~\bibnamefont {Chen}}, \bibinfo {author} {\bibfnamefont {W.}~\bibnamefont {Wu}}, \bibinfo {author} {\bibfnamefont {Z.-M.}\ \bibnamefont {Yu}}, \bibinfo {author} {\bibfnamefont {Z.}~\bibnamefont {Chen}}, \bibinfo {author} {\bibfnamefont {Y.~X.}\ \bibnamefont {Zhao}}, \bibinfo {author} {\bibfnamefont {X.-L.}\ \bibnamefont {Sheng}},\ and\ \bibinfo {author} {\bibfnamefont {S.~A.}\ \bibnamefont {Yang}},\ }\bibfield  {title} {\bibinfo {title} {Graphyne as a second-order and real {Chern} topological insulator in two dimensions},\ }\href@noop {} {\bibfield  {journal} {\bibinfo  {journal} {Phys. Rev. B}\ }\textbf {\bibinfo {volume} {104}},\ \bibinfo {pages} {085205} (\bibinfo {year} {2021})}\BibitemShut {NoStop}%
\bibitem [{\citenamefont {Chen}\ \emph {et~al.}(2022)\citenamefont {Chen}, \citenamefont {Zeng}, \citenamefont {Chen}, \citenamefont {Zhao}, \citenamefont {Sheng},\ and\ \citenamefont {Yang}}]{Chen2022Second-PRL}%
  \BibitemOpen
  \bibfield  {author} {\bibinfo {author} {\bibfnamefont {C.}~\bibnamefont {Chen}}, \bibinfo {author} {\bibfnamefont {X.-T.}\ \bibnamefont {Zeng}}, \bibinfo {author} {\bibfnamefont {Z.}~\bibnamefont {Chen}}, \bibinfo {author} {\bibfnamefont {Y.~X.}\ \bibnamefont {Zhao}}, \bibinfo {author} {\bibfnamefont {X.-L.}\ \bibnamefont {Sheng}},\ and\ \bibinfo {author} {\bibfnamefont {S.~A.}\ \bibnamefont {Yang}},\ }\bibfield  {title} {\bibinfo {title} {Second-order real nodal-line semimetal in three-dimensional graphdiyne},\ }\href@noop {} {\bibfield  {journal} {\bibinfo  {journal} {Phys. Rev. Lett.}\ }\textbf {\bibinfo {volume} {128}},\ \bibinfo {pages} {026405} (\bibinfo {year} {2022})}\BibitemShut {NoStop}%
\bibitem [{\citenamefont {Benalcazar}\ \emph {et~al.}(2017)\citenamefont {Benalcazar}, \citenamefont {Bernevig},\ and\ \citenamefont {Hughes}}]{Benalcazar2017Quantized-S}%
  \BibitemOpen
  \bibfield  {author} {\bibinfo {author} {\bibfnamefont {W.~A.}\ \bibnamefont {Benalcazar}}, \bibinfo {author} {\bibfnamefont {B.~A.}\ \bibnamefont {Bernevig}},\ and\ \bibinfo {author} {\bibfnamefont {T.~L.}\ \bibnamefont {Hughes}},\ }\bibfield  {title} {\bibinfo {title} {Quantized electric multipole insulators},\ }\href@noop {} {\bibfield  {journal} {\bibinfo  {journal} {Science}\ }\textbf {\bibinfo {volume} {357}},\ \bibinfo {pages} {61} (\bibinfo {year} {2017})}\BibitemShut {NoStop}%
\bibitem [{\citenamefont {Song}\ \emph {et~al.}(2017)\citenamefont {Song}, \citenamefont {Fang},\ and\ \citenamefont {Fang}}]{Song2017d-PRL}%
  \BibitemOpen
  \bibfield  {author} {\bibinfo {author} {\bibfnamefont {Z.}~\bibnamefont {Song}}, \bibinfo {author} {\bibfnamefont {Z.}~\bibnamefont {Fang}},\ and\ \bibinfo {author} {\bibfnamefont {C.}~\bibnamefont {Fang}},\ }\bibfield  {title} {\bibinfo {title} {$(d\ensuremath{-}2)$-dimensional edge states of rotation symmetry protected topological states},\ }\href@noop {} {\bibfield  {journal} {\bibinfo  {journal} {Phys. Rev. Lett.}\ }\textbf {\bibinfo {volume} {119}},\ \bibinfo {pages} {246402} (\bibinfo {year} {2017})}\BibitemShut {NoStop}%
\bibitem [{\citenamefont {Langbehn}\ \emph {et~al.}(2017)\citenamefont {Langbehn}, \citenamefont {Peng}, \citenamefont {Trifunovic}, \citenamefont {von Oppen},\ and\ \citenamefont {Brouwer}}]{Langbehn2017Reflection-PRL}%
  \BibitemOpen
  \bibfield  {author} {\bibinfo {author} {\bibfnamefont {J.}~\bibnamefont {Langbehn}}, \bibinfo {author} {\bibfnamefont {Y.}~\bibnamefont {Peng}}, \bibinfo {author} {\bibfnamefont {L.}~\bibnamefont {Trifunovic}}, \bibinfo {author} {\bibfnamefont {F.}~\bibnamefont {von Oppen}},\ and\ \bibinfo {author} {\bibfnamefont {P.~W.}\ \bibnamefont {Brouwer}},\ }\bibfield  {title} {\bibinfo {title} {Reflection-symmetric second-order topological insulators and superconductors},\ }\href@noop {} {\bibfield  {journal} {\bibinfo  {journal} {Phys. Rev. Lett.}\ }\textbf {\bibinfo {volume} {119}},\ \bibinfo {pages} {246401} (\bibinfo {year} {2017})}\BibitemShut {NoStop}%
\bibitem [{\citenamefont {Schindler}\ \emph {et~al.}(2018)\citenamefont {Schindler}, \citenamefont {Cook}, \citenamefont {Vergniory}, \citenamefont {Wang}, \citenamefont {Parkin}, \citenamefont {Bernevig},\ and\ \citenamefont {Neupert}}]{Schindler2018High-SA}%
  \BibitemOpen
  \bibfield  {author} {\bibinfo {author} {\bibfnamefont {F.}~\bibnamefont {Schindler}}, \bibinfo {author} {\bibfnamefont {A.~M.}\ \bibnamefont {Cook}}, \bibinfo {author} {\bibfnamefont {M.~G.}\ \bibnamefont {Vergniory}}, \bibinfo {author} {\bibfnamefont {Z.}~\bibnamefont {Wang}}, \bibinfo {author} {\bibfnamefont {S.~S.~P.}\ \bibnamefont {Parkin}}, \bibinfo {author} {\bibfnamefont {B.~A.}\ \bibnamefont {Bernevig}},\ and\ \bibinfo {author} {\bibfnamefont {T.}~\bibnamefont {Neupert}},\ }\bibfield  {title} {\bibinfo {title} {Higher-order topological insulators},\ }\href@noop {} {\bibfield  {journal} {\bibinfo  {journal} {Sci. Adv.}\ }\textbf {\bibinfo {volume} {4}},\ \bibinfo {pages} {eaat0346} (\bibinfo {year} {2018})}\BibitemShut {NoStop}%
\bibitem [{\citenamefont {Sheng}\ \emph {et~al.}(2019)\citenamefont {Sheng}, \citenamefont {Chen}, \citenamefont {Liu}, \citenamefont {Chen}, \citenamefont {Yu}, \citenamefont {Zhao},\ and\ \citenamefont {Yang}}]{Sheng2019Two-PRL}%
  \BibitemOpen
  \bibfield  {author} {\bibinfo {author} {\bibfnamefont {X.-L.}\ \bibnamefont {Sheng}}, \bibinfo {author} {\bibfnamefont {C.}~\bibnamefont {Chen}}, \bibinfo {author} {\bibfnamefont {H.}~\bibnamefont {Liu}}, \bibinfo {author} {\bibfnamefont {Z.}~\bibnamefont {Chen}}, \bibinfo {author} {\bibfnamefont {Z.-M.}\ \bibnamefont {Yu}}, \bibinfo {author} {\bibfnamefont {Y.~X.}\ \bibnamefont {Zhao}},\ and\ \bibinfo {author} {\bibfnamefont {S.~A.}\ \bibnamefont {Yang}},\ }\bibfield  {title} {\bibinfo {title} {Two-dimensional second-order topological insulator in graphdiyne},\ }\href@noop {} {\bibfield  {journal} {\bibinfo  {journal} {Phys. Rev. Lett.}\ }\textbf {\bibinfo {volume} {123}},\ \bibinfo {pages} {256402} (\bibinfo {year} {2019})}\BibitemShut {NoStop}%
\bibitem [{\citenamefont {Wang}\ \emph {et~al.}(2019)\citenamefont {Wang}, \citenamefont {Wieder}, \citenamefont {Li}, \citenamefont {Yan},\ and\ \citenamefont {Bernevig}}]{Wang2019Higher-PRL}%
  \BibitemOpen
  \bibfield  {author} {\bibinfo {author} {\bibfnamefont {Z.}~\bibnamefont {Wang}}, \bibinfo {author} {\bibfnamefont {B.~J.}\ \bibnamefont {Wieder}}, \bibinfo {author} {\bibfnamefont {J.}~\bibnamefont {Li}}, \bibinfo {author} {\bibfnamefont {B.}~\bibnamefont {Yan}},\ and\ \bibinfo {author} {\bibfnamefont {B.~A.}\ \bibnamefont {Bernevig}},\ }\bibfield  {title} {\bibinfo {title} {Higher-order topology, monopole nodal lines, and the origin of large {Fermi} arcs in transition metal dichalcogenides {XTe$_2$} ({X} = {Mo}, {W})},\ }\href@noop {} {\bibfield  {journal} {\bibinfo  {journal} {Phys. Rev. Lett.}\ }\textbf {\bibinfo {volume} {123}},\ \bibinfo {pages} {186401} (\bibinfo {year} {2019})}\BibitemShut {NoStop}%
\bibitem [{\citenamefont {Lee}\ \emph {et~al.}(2020)\citenamefont {Lee}, \citenamefont {Kim}, \citenamefont {Ahn},\ and\ \citenamefont {Yang}}]{Lee2020Graphdiyne-nQM}%
  \BibitemOpen
  \bibfield  {author} {\bibinfo {author} {\bibfnamefont {E.}~\bibnamefont {Lee}}, \bibinfo {author} {\bibfnamefont {R.}~\bibnamefont {Kim}}, \bibinfo {author} {\bibfnamefont {J.}~\bibnamefont {Ahn}},\ and\ \bibinfo {author} {\bibfnamefont {B.-J.}\ \bibnamefont {Yang}},\ }\bibfield  {title} {\bibinfo {title} {Two-dimensional higher-order topology in monolayer graphdiyne},\ }\href@noop {} {\bibfield  {journal} {\bibinfo  {journal} {npj Quantum Mater.}\ }\textbf {\bibinfo {volume} {5}},\ \bibinfo {pages} {1} (\bibinfo {year} {2020})}\BibitemShut {NoStop}%
\bibitem [{\citenamefont {Wang}\ \emph {et~al.}(2024{\natexlab{a}})\citenamefont {Wang}, \citenamefont {Zhang}, \citenamefont {Zhang}, \citenamefont {Cheng}, \citenamefont {Wang}, \citenamefont {Qian}, \citenamefont {Cheng}, \citenamefont {Zhang},\ and\ \citenamefont {Wang}}]{Wang20243D-AFM}%
  \BibitemOpen
  \bibfield  {author} {\bibinfo {author} {\bibfnamefont {J.}~\bibnamefont {Wang}}, \bibinfo {author} {\bibfnamefont {T.-T.}\ \bibnamefont {Zhang}}, \bibinfo {author} {\bibfnamefont {Q.}~\bibnamefont {Zhang}}, \bibinfo {author} {\bibfnamefont {X.}~\bibnamefont {Cheng}}, \bibinfo {author} {\bibfnamefont {W.}~\bibnamefont {Wang}}, \bibinfo {author} {\bibfnamefont {S.}~\bibnamefont {Qian}}, \bibinfo {author} {\bibfnamefont {Z.}~\bibnamefont {Cheng}}, \bibinfo {author} {\bibfnamefont {G.}~\bibnamefont {Zhang}},\ and\ \bibinfo {author} {\bibfnamefont {X.}~\bibnamefont {Wang}},\ }\bibfield  {title} {\bibinfo {title} {{3D} carbon allotropes: Topological quantum materials with obstructed atomic insulating phases, multiple bulk-boundary correspondences, and real topology},\ }\href@noop {} {\bibfield  {journal} {\bibinfo  {journal} {Adv. Funct. Mater.}\ }\textbf {\bibinfo {volume} {34}},\ \bibinfo {pages} {2316079} (\bibinfo {year} {2024}{\natexlab{a}})}\BibitemShut {NoStop}%
\bibitem [{\citenamefont {Wang}\ \emph {et~al.}(2024{\natexlab{b}})\citenamefont {Wang}, \citenamefont {Bai}, \citenamefont {Wang}, \citenamefont {Cheng}, \citenamefont {Qian}, \citenamefont {Wang}, \citenamefont {Zhang}, \citenamefont {Yu},\ and\ \citenamefont {Yao}}]{Wang2024Real-AM}%
  \BibitemOpen
  \bibfield  {author} {\bibinfo {author} {\bibfnamefont {X.}~\bibnamefont {Wang}}, \bibinfo {author} {\bibfnamefont {J.}~\bibnamefont {Bai}}, \bibinfo {author} {\bibfnamefont {J.}~\bibnamefont {Wang}}, \bibinfo {author} {\bibfnamefont {Z.}~\bibnamefont {Cheng}}, \bibinfo {author} {\bibfnamefont {S.}~\bibnamefont {Qian}}, \bibinfo {author} {\bibfnamefont {W.}~\bibnamefont {Wang}}, \bibinfo {author} {\bibfnamefont {G.}~\bibnamefont {Zhang}}, \bibinfo {author} {\bibfnamefont {Z.-M.}\ \bibnamefont {Yu}},\ and\ \bibinfo {author} {\bibfnamefont {Y.}~\bibnamefont {Yao}},\ }\bibfield  {title} {\bibinfo {title} {Real topological phonons in {3D} carbon allotropes},\ }\href@noop {} {\bibfield  {journal} {\bibinfo  {journal} {Adv. Mater.}\ }\textbf {\bibinfo {volume} {36}},\ \bibinfo {pages} {2407437} (\bibinfo {year} {2024}{\natexlab{b}})}\BibitemShut {NoStop}%
\bibitem [{\citenamefont {Han}\ \emph {et~al.}(2024)\citenamefont {Han}, \citenamefont {Liu}, \citenamefont {Cui}, \citenamefont {Liu},\ and\ \citenamefont {Yu}}]{Han2024Crossed-PRB}%
  \BibitemOpen
  \bibfield  {author} {\bibinfo {author} {\bibfnamefont {Y.}~\bibnamefont {Han}}, \bibinfo {author} {\bibfnamefont {Y.}~\bibnamefont {Liu}}, \bibinfo {author} {\bibfnamefont {C.}~\bibnamefont {Cui}}, \bibinfo {author} {\bibfnamefont {C.-C.}\ \bibnamefont {Liu}},\ and\ \bibinfo {author} {\bibfnamefont {Z.-M.}\ \bibnamefont {Yu}},\ }\bibfield  {title} {\bibinfo {title} {Crossed real nodal-line phonons in gold monobromide},\ }\href@noop {} {\bibfield  {journal} {\bibinfo  {journal} {Phys. Rev. B}\ }\textbf {\bibinfo {volume} {110}},\ \bibinfo {pages} {184303} (\bibinfo {year} {2024})}\BibitemShut {NoStop}%
\bibitem [{\citenamefont {Xue}\ \emph {et~al.}(2023)\citenamefont {Xue}, \citenamefont {Chen}, \citenamefont {Cheng}, \citenamefont {Dai}, \citenamefont {Long}, \citenamefont {Zhao},\ and\ \citenamefont {Zhang}}]{Xue2023Stiefel-NC}%
  \BibitemOpen
  \bibfield  {author} {\bibinfo {author} {\bibfnamefont {H.}~\bibnamefont {Xue}}, \bibinfo {author} {\bibfnamefont {Z.~Y.}\ \bibnamefont {Chen}}, \bibinfo {author} {\bibfnamefont {Z.}~\bibnamefont {Cheng}}, \bibinfo {author} {\bibfnamefont {J.~X.}\ \bibnamefont {Dai}}, \bibinfo {author} {\bibfnamefont {Y.}~\bibnamefont {Long}}, \bibinfo {author} {\bibfnamefont {Y.~X.}\ \bibnamefont {Zhao}},\ and\ \bibinfo {author} {\bibfnamefont {B.}~\bibnamefont {Zhang}},\ }\bibfield  {title} {\bibinfo {title} {{Stiefel-Whitney} topological charges in a three-dimensional acoustic nodal-line crystal},\ }\href@noop {} {\bibfield  {journal} {\bibinfo  {journal} {Nat. Commun.}\ }\textbf {\bibinfo {volume} {14}},\ \bibinfo {pages} {4563} (\bibinfo {year} {2023})}\BibitemShut {NoStop}%
\bibitem [{\citenamefont {Xiang}\ \emph {et~al.}(2024)\citenamefont {Xiang}, \citenamefont {Peng}, \citenamefont {Gao}, \citenamefont {Wu}, \citenamefont {Wu}, \citenamefont {Chen}, \citenamefont {Ni},\ and\ \citenamefont {Zhu}}]{Xiang2024Demonstration-PRL}%
  \BibitemOpen
  \bibfield  {author} {\bibinfo {author} {\bibfnamefont {X.}~\bibnamefont {Xiang}}, \bibinfo {author} {\bibfnamefont {Y.-G.}\ \bibnamefont {Peng}}, \bibinfo {author} {\bibfnamefont {F.}~\bibnamefont {Gao}}, \bibinfo {author} {\bibfnamefont {X.}~\bibnamefont {Wu}}, \bibinfo {author} {\bibfnamefont {P.}~\bibnamefont {Wu}}, \bibinfo {author} {\bibfnamefont {Z.}~\bibnamefont {Chen}}, \bibinfo {author} {\bibfnamefont {X.}~\bibnamefont {Ni}},\ and\ \bibinfo {author} {\bibfnamefont {X.-F.}\ \bibnamefont {Zhu}},\ }\bibfield  {title} {\bibinfo {title} {Demonstration of acoustic higher-order topological {Stiefel-Whitney} semimetal},\ }\href@noop {} {\bibfield  {journal} {\bibinfo  {journal} {Phys. Rev. Lett.}\ }\textbf {\bibinfo {volume} {132}},\ \bibinfo {pages} {197202} (\bibinfo {year} {2024})}\BibitemShut {NoStop}%
\bibitem [{\citenamefont {Ma}\ \emph {et~al.}(2024)\citenamefont {Ma}, \citenamefont {Pu}, \citenamefont {Ye}, \citenamefont {Lu}, \citenamefont {Huang}, \citenamefont {Ke}, \citenamefont {He}, \citenamefont {Deng},\ and\ \citenamefont {Liu}}]{Ma2024Observation-PRL}%
  \BibitemOpen
  \bibfield  {author} {\bibinfo {author} {\bibfnamefont {Q.}~\bibnamefont {Ma}}, \bibinfo {author} {\bibfnamefont {Z.}~\bibnamefont {Pu}}, \bibinfo {author} {\bibfnamefont {L.}~\bibnamefont {Ye}}, \bibinfo {author} {\bibfnamefont {J.}~\bibnamefont {Lu}}, \bibinfo {author} {\bibfnamefont {X.}~\bibnamefont {Huang}}, \bibinfo {author} {\bibfnamefont {M.}~\bibnamefont {Ke}}, \bibinfo {author} {\bibfnamefont {H.}~\bibnamefont {He}}, \bibinfo {author} {\bibfnamefont {W.}~\bibnamefont {Deng}},\ and\ \bibinfo {author} {\bibfnamefont {Z.}~\bibnamefont {Liu}},\ }\bibfield  {title} {\bibinfo {title} {Observation of higher-order nodal-line semimetal in phononic crystals},\ }\href@noop {} {\bibfield  {journal} {\bibinfo  {journal} {Phys. Rev. Lett.}\ }\textbf {\bibinfo {volume} {132}},\ \bibinfo {pages} {066601} (\bibinfo {year} {2024})}\BibitemShut {NoStop}%
\bibitem [{\citenamefont {Li}\ \emph {et~al.}(2025)\citenamefont {Li}, \citenamefont {Qian},\ and\ \citenamefont {Liu}}]{Li2025General-PRB}%
  \BibitemOpen
  \bibfield  {author} {\bibinfo {author} {\bibfnamefont {Y.}~\bibnamefont {Li}}, \bibinfo {author} {\bibfnamefont {S.}~\bibnamefont {Qian}},\ and\ \bibinfo {author} {\bibfnamefont {C.-C.}\ \bibnamefont {Liu}},\ }\bibfield  {title} {\bibinfo {title} {General construction of three-dimensional {$\mathbb{Z}_2$} monopole charge nodal line semimetals and prediction of abundant candidate materials},\ }\href@noop {} {\bibfield  {journal} {\bibinfo  {journal} {Phys. Rev. B}\ }\textbf {\bibinfo {volume} {111}},\ \bibinfo {pages} {125101} (\bibinfo {year} {2025})}\BibitemShut {NoStop}%
\bibitem [{\citenamefont {Li}\ \emph {et~al.}(2017)\citenamefont {Li}, \citenamefont {Yu}, \citenamefont {Liu}, \citenamefont {Guan}, \citenamefont {Wang}, \citenamefont {Zhang}, \citenamefont {Yao},\ and\ \citenamefont {Yang}}]{Li2017Type-PRB}%
  \BibitemOpen
  \bibfield  {author} {\bibinfo {author} {\bibfnamefont {S.}~\bibnamefont {Li}}, \bibinfo {author} {\bibfnamefont {Z.-M.}\ \bibnamefont {Yu}}, \bibinfo {author} {\bibfnamefont {Y.}~\bibnamefont {Liu}}, \bibinfo {author} {\bibfnamefont {S.}~\bibnamefont {Guan}}, \bibinfo {author} {\bibfnamefont {S.-S.}\ \bibnamefont {Wang}}, \bibinfo {author} {\bibfnamefont {X.}~\bibnamefont {Zhang}}, \bibinfo {author} {\bibfnamefont {Y.}~\bibnamefont {Yao}},\ and\ \bibinfo {author} {\bibfnamefont {S.~A.}\ \bibnamefont {Yang}},\ }\bibfield  {title} {\bibinfo {title} {{Type-II} nodal loops: Theory and material realization},\ }\href@noop {} {\bibfield  {journal} {\bibinfo  {journal} {Phys. Rev. B}\ }\textbf {\bibinfo {volume} {96}},\ \bibinfo {pages} {081106} (\bibinfo {year} {2017})}\BibitemShut {NoStop}%
\bibitem [{Sup()}]{SupplementaryMaterial}%
  \BibitemOpen
  \href@noop {} {}\bibinfo {note} {See Supplemental Material at [URL will be inserted by publisher] for details of the tight-binding model construction, Wilson-loop calculations of the real Chern number, the evolution of real nodal lines in IGN, and first-principles calculations; this material includes Refs.~\mbox{[62--72]}.}\BibitemShut {Stop}%
\bibitem [{\citenamefont {Yue}\ \emph {et~al.}(2024)\citenamefont {Yue}, \citenamefont {Liu}, \citenamefont {Yang},\ and\ \citenamefont {Zhao}}]{Yue2024Stability-PRB}%
  \BibitemOpen
  \bibfield  {author} {\bibinfo {author} {\bibfnamefont {S.~J.}\ \bibnamefont {Yue}}, \bibinfo {author} {\bibfnamefont {Q.}~\bibnamefont {Liu}}, \bibinfo {author} {\bibfnamefont {S.~A.}\ \bibnamefont {Yang}},\ and\ \bibinfo {author} {\bibfnamefont {Y.~X.}\ \bibnamefont {Zhao}},\ }\bibfield  {title} {\bibinfo {title} {Stability and noncentered {PT} symmetry of real topological phases},\ }\href@noop {} {\bibfield  {journal} {\bibinfo  {journal} {Phys. Rev. B}\ }\textbf {\bibinfo {volume} {109}},\ \bibinfo {pages} {195116} (\bibinfo {year} {2024})}\BibitemShut {NoStop}%
\bibitem [{\citenamefont {Zhu}\ \emph {et~al.}(2022)\citenamefont {Zhu}, \citenamefont {Wu}, \citenamefont {Zhao}, \citenamefont {Chen}, \citenamefont {Wang}, \citenamefont {Sheng}, \citenamefont {Zhang}, \citenamefont {Zhao},\ and\ \citenamefont {Yang}}]{Zhu2022Phononic-PRB}%
  \BibitemOpen
  \bibfield  {author} {\bibinfo {author} {\bibfnamefont {J.}~\bibnamefont {Zhu}}, \bibinfo {author} {\bibfnamefont {W.}~\bibnamefont {Wu}}, \bibinfo {author} {\bibfnamefont {J.}~\bibnamefont {Zhao}}, \bibinfo {author} {\bibfnamefont {C.}~\bibnamefont {Chen}}, \bibinfo {author} {\bibfnamefont {Q.}~\bibnamefont {Wang}}, \bibinfo {author} {\bibfnamefont {X.-L.}\ \bibnamefont {Sheng}}, \bibinfo {author} {\bibfnamefont {L.}~\bibnamefont {Zhang}}, \bibinfo {author} {\bibfnamefont {Y.~X.}\ \bibnamefont {Zhao}},\ and\ \bibinfo {author} {\bibfnamefont {S.~A.}\ \bibnamefont {Yang}},\ }\bibfield  {title} {\bibinfo {title} {Phononic real {Chern} insulator with protected corner modes in graphynes},\ }\href@noop {} {\bibfield  {journal} {\bibinfo  {journal} {Phys. Rev. B}\ }\textbf {\bibinfo {volume} {105}},\ \bibinfo {pages} {085123} (\bibinfo {year} {2022})}\BibitemShut {NoStop}%
\bibitem [{\citenamefont {Yang}\ \emph {et~al.}(2023)\citenamefont {Yang}, \citenamefont {Gao}, \citenamefont {Hao}, \citenamefont {Zhang}, \citenamefont {Tan}, \citenamefont {Wang}, \citenamefont {Cheng},\ and\ \citenamefont {Wu}}]{Yang2023Cladded-PRB}%
  \BibitemOpen
  \bibfield  {author} {\bibinfo {author} {\bibfnamefont {T.}~\bibnamefont {Yang}}, \bibinfo {author} {\bibfnamefont {Y.}~\bibnamefont {Gao}}, \bibinfo {author} {\bibfnamefont {L.}~\bibnamefont {Hao}}, \bibinfo {author} {\bibfnamefont {H.}~\bibnamefont {Zhang}}, \bibinfo {author} {\bibfnamefont {X.}~\bibnamefont {Tan}}, \bibinfo {author} {\bibfnamefont {P.}~\bibnamefont {Wang}}, \bibinfo {author} {\bibfnamefont {Z.}~\bibnamefont {Cheng}},\ and\ \bibinfo {author} {\bibfnamefont {W.}~\bibnamefont {Wu}},\ }\bibfield  {title} {\bibinfo {title} {Cladded phononic nodal frame state in biatomic alkali-metal sulfides},\ }\href@noop {} {\bibfield  {journal} {\bibinfo  {journal} {Phys. Rev. B}\ }\textbf {\bibinfo {volume} {108}},\ \bibinfo {pages} {134310} (\bibinfo {year} {2023})}\BibitemShut {NoStop}%
\bibitem [{\citenamefont {Xiao}\ and\ \citenamefont {Fan}(2017)}]{Xiao2017Topologically-arXiv}%
  \BibitemOpen
  \bibfield  {author} {\bibinfo {author} {\bibfnamefont {M.}~\bibnamefont {Xiao}}\ and\ \bibinfo {author} {\bibfnamefont {S.}~\bibnamefont {Fan}},\ }\bibfield  {title} {\bibinfo {title} {Topologically charged nodal surface},\ }\href@noop {} {\bibfield  {journal} {\bibinfo  {journal} {arXiv:1709.02363}\ } (\bibinfo {year} {2017})}\BibitemShut {NoStop}%
\bibitem [{\citenamefont {Kim}\ \emph {et~al.}(2019)\citenamefont {Kim}, \citenamefont {Lee}, \citenamefont {Lee},\ and\ \citenamefont {Rho}}]{Kim2019Topologically-PRB}%
  \BibitemOpen
  \bibfield  {author} {\bibinfo {author} {\bibfnamefont {M.}~\bibnamefont {Kim}}, \bibinfo {author} {\bibfnamefont {D.}~\bibnamefont {Lee}}, \bibinfo {author} {\bibfnamefont {D.}~\bibnamefont {Lee}},\ and\ \bibinfo {author} {\bibfnamefont {J.}~\bibnamefont {Rho}},\ }\bibfield  {title} {\bibinfo {title} {Topologically nontrivial photonic nodal surface in a photonic metamaterial},\ }\href@noop {} {\bibfield  {journal} {\bibinfo  {journal} {Phys. Rev. B}\ }\textbf {\bibinfo {volume} {99}},\ \bibinfo {pages} {235423} (\bibinfo {year} {2019})}\BibitemShut {NoStop}%
\bibitem [{\citenamefont {Narayan}(2016)}]{Narayan2016Tunable-PRB}%
  \BibitemOpen
  \bibfield  {author} {\bibinfo {author} {\bibfnamefont {A.}~\bibnamefont {Narayan}},\ }\bibfield  {title} {\bibinfo {title} {Tunable point nodes from line-node semimetals via application of light},\ }\href@noop {} {\bibfield  {journal} {\bibinfo  {journal} {Phys. Rev. B}\ }\textbf {\bibinfo {volume} {94}},\ \bibinfo {pages} {041409(R)} (\bibinfo {year} {2016})}\BibitemShut {NoStop}%
\bibitem [{\citenamefont {Ezawa}(2017)}]{Ezawa2017Photoinduced-PRB}%
  \BibitemOpen
  \bibfield  {author} {\bibinfo {author} {\bibfnamefont {M.}~\bibnamefont {Ezawa}},\ }\bibfield  {title} {\bibinfo {title} {Photoinduced topological phase transition from a crossing-line nodal semimetal to a multiple-{Weyl} semimetal},\ }\href@noop {} {\bibfield  {journal} {\bibinfo  {journal} {Phys. Rev. B}\ }\textbf {\bibinfo {volume} {96}},\ \bibinfo {pages} {041205(R)} (\bibinfo {year} {2017})}\BibitemShut {NoStop}%
\bibitem [{\citenamefont {Yan}\ and\ \citenamefont {Wang}(2017)}]{Yan2017Floquet-PRB}%
  \BibitemOpen
  \bibfield  {author} {\bibinfo {author} {\bibfnamefont {Z.}~\bibnamefont {Yan}}\ and\ \bibinfo {author} {\bibfnamefont {Z.}~\bibnamefont {Wang}},\ }\bibfield  {title} {\bibinfo {title} {{Floquet} multi-{Weyl} points in crossing-nodal-line semimetals},\ }\href@noop {} {\bibfield  {journal} {\bibinfo  {journal} {Phys. Rev. B}\ }\textbf {\bibinfo {volume} {96}},\ \bibinfo {pages} {041206(R)} (\bibinfo {year} {2017})}\BibitemShut {NoStop}%
\bibitem [{\citenamefont {Du}\ \emph {et~al.}(2022)\citenamefont {Du}, \citenamefont {Chen}, \citenamefont {Wang},\ and\ \citenamefont {Xu}}]{Du2022Weyl-PRB}%
  \BibitemOpen
  \bibfield  {author} {\bibinfo {author} {\bibfnamefont {X.-L.}\ \bibnamefont {Du}}, \bibinfo {author} {\bibfnamefont {R.}~\bibnamefont {Chen}}, \bibinfo {author} {\bibfnamefont {R.}~\bibnamefont {Wang}},\ and\ \bibinfo {author} {\bibfnamefont {D.-H.}\ \bibnamefont {Xu}},\ }\bibfield  {title} {\bibinfo {title} {Weyl nodes with higher-order topology in an optically driven nodal-line semimetal},\ }\href@noop {} {\bibfield  {journal} {\bibinfo  {journal} {Phys. Rev. B}\ }\textbf {\bibinfo {volume} {105}},\ \bibinfo {pages} {L081102} (\bibinfo {year} {2022})}\BibitemShut {NoStop}%
\bibitem [{\citenamefont {Liu}\ \emph {et~al.}(2025)\citenamefont {Liu}, \citenamefont {Cui}, \citenamefont {Li}, \citenamefont {Li}, \citenamefont {Xu},\ and\ \citenamefont {Yu}}]{Liu2025Floquet-PRB}%
  \BibitemOpen
  \bibfield  {author} {\bibinfo {author} {\bibfnamefont {P.}~\bibnamefont {Liu}}, \bibinfo {author} {\bibfnamefont {C.}~\bibnamefont {Cui}}, \bibinfo {author} {\bibfnamefont {L.}~\bibnamefont {Li}}, \bibinfo {author} {\bibfnamefont {R.}~\bibnamefont {Li}}, \bibinfo {author} {\bibfnamefont {D.-H.}\ \bibnamefont {Xu}},\ and\ \bibinfo {author} {\bibfnamefont {Z.-M.}\ \bibnamefont {Yu}},\ }\bibfield  {title} {\bibinfo {title} {{Floquet} control of topological phases and {Hall} effects in {$\mathbb{Z}_2$} nodal line semimetals},\ }\href@noop {} {\bibfield  {journal} {\bibinfo  {journal} {Phys. Rev. B}\ }\textbf {\bibinfo {volume} {111}},\ \bibinfo {pages} {235105} (\bibinfo {year} {2025})}\BibitemShut {NoStop}%
\bibitem [{\citenamefont {Zhang}\ \emph {et~al.}(2022)\citenamefont {Zhang}, \citenamefont {Yu}, \citenamefont {Liu},\ and\ \citenamefont {Yao}}]{Zhang2022MagneticTB-CPC}%
  \BibitemOpen
  \bibfield  {author} {\bibinfo {author} {\bibfnamefont {Z.}~\bibnamefont {Zhang}}, \bibinfo {author} {\bibfnamefont {Z.-M.}\ \bibnamefont {Yu}}, \bibinfo {author} {\bibfnamefont {G.-B.}\ \bibnamefont {Liu}},\ and\ \bibinfo {author} {\bibfnamefont {Y.}~\bibnamefont {Yao}},\ }\bibfield  {title} {\bibinfo {title} {{\mbox{MagneticTB}}: A package for tight-binding model of magnetic and non-magnetic materials},\ }\href@noop {} {\bibfield  {journal} {\bibinfo  {journal} {Comput. Phys. Commun.}\ }\textbf {\bibinfo {volume} {270}},\ \bibinfo {pages} {108153} (\bibinfo {year} {2022})}\BibitemShut {NoStop}%
\bibitem [{\citenamefont {Kresse}\ and\ \citenamefont {Hafner}(1993)}]{Kresse1993Ab-PRB}%
  \BibitemOpen
  \bibfield  {author} {\bibinfo {author} {\bibfnamefont {G.}~\bibnamefont {Kresse}}\ and\ \bibinfo {author} {\bibfnamefont {J.}~\bibnamefont {Hafner}},\ }\bibfield  {title} {\bibinfo {title} {Ab initio molecular dynamics for liquid metals},\ }\href@noop {} {\bibfield  {journal} {\bibinfo  {journal} {Phys. Rev. B}\ }\textbf {\bibinfo {volume} {47}},\ \bibinfo {pages} {558} (\bibinfo {year} {1993})}\BibitemShut {NoStop}%
\bibitem [{\citenamefont {Kresse}\ and\ \citenamefont {Furthm{\"u}ller}(1996)}]{Kresse1996Efficient-PRB}%
  \BibitemOpen
  \bibfield  {author} {\bibinfo {author} {\bibfnamefont {G.}~\bibnamefont {Kresse}}\ and\ \bibinfo {author} {\bibfnamefont {J.}~\bibnamefont {Furthm{\"u}ller}},\ }\bibfield  {title} {\bibinfo {title} {Efficient iterative schemes for ab initio total-energy calculations using a plane-wave basis set},\ }\href@noop {} {\bibfield  {journal} {\bibinfo  {journal} {Phys. Rev. B}\ }\textbf {\bibinfo {volume} {54}},\ \bibinfo {pages} {11169} (\bibinfo {year} {1996})}\BibitemShut {NoStop}%
\bibitem [{\citenamefont {Bl{\"o}chl}(1994)}]{Bloechl1994Projector-PRB}%
  \BibitemOpen
  \bibfield  {author} {\bibinfo {author} {\bibfnamefont {P.~E.}\ \bibnamefont {Bl{\"o}chl}},\ }\bibfield  {title} {\bibinfo {title} {Projector augmented-wave method},\ }\href@noop {} {\bibfield  {journal} {\bibinfo  {journal} {Phys. Rev. B}\ }\textbf {\bibinfo {volume} {50}},\ \bibinfo {pages} {17953} (\bibinfo {year} {1994})}\BibitemShut {NoStop}%
\bibitem [{\citenamefont {Perdew}\ \emph {et~al.}(1996)\citenamefont {Perdew}, \citenamefont {Burke},\ and\ \citenamefont {Ernzerhof}}]{Perdew1996Generalized-PRL}%
  \BibitemOpen
  \bibfield  {author} {\bibinfo {author} {\bibfnamefont {J.~P.}\ \bibnamefont {Perdew}}, \bibinfo {author} {\bibfnamefont {K.}~\bibnamefont {Burke}},\ and\ \bibinfo {author} {\bibfnamefont {M.}~\bibnamefont {Ernzerhof}},\ }\bibfield  {title} {\bibinfo {title} {Generalized gradient approximation made simple},\ }\href@noop {} {\bibfield  {journal} {\bibinfo  {journal} {Phys. Rev. Lett.}\ }\textbf {\bibinfo {volume} {77}},\ \bibinfo {pages} {3865} (\bibinfo {year} {1996})}\BibitemShut {NoStop}%
\bibitem [{\citenamefont {Marzari}\ and\ \citenamefont {Vanderbilt}(1997)}]{Marzari1997Maximally-PRB}%
  \BibitemOpen
  \bibfield  {author} {\bibinfo {author} {\bibfnamefont {N.}~\bibnamefont {Marzari}}\ and\ \bibinfo {author} {\bibfnamefont {D.}~\bibnamefont {Vanderbilt}},\ }\bibfield  {title} {\bibinfo {title} {Maximally localized generalized {Wannier} functions for composite energy bands},\ }\href@noop {} {\bibfield  {journal} {\bibinfo  {journal} {Phys. Rev. B}\ }\textbf {\bibinfo {volume} {56}},\ \bibinfo {pages} {12847} (\bibinfo {year} {1997})}\BibitemShut {NoStop}%
\bibitem [{\citenamefont {Souza}\ \emph {et~al.}(2001)\citenamefont {Souza}, \citenamefont {Marzari},\ and\ \citenamefont {Vanderbilt}}]{Souza2001Maximally-PRB}%
  \BibitemOpen
  \bibfield  {author} {\bibinfo {author} {\bibfnamefont {I.}~\bibnamefont {Souza}}, \bibinfo {author} {\bibfnamefont {N.}~\bibnamefont {Marzari}},\ and\ \bibinfo {author} {\bibfnamefont {D.}~\bibnamefont {Vanderbilt}},\ }\bibfield  {title} {\bibinfo {title} {Maximally localized {Wannier} functions for entangled energy bands},\ }\href@noop {} {\bibfield  {journal} {\bibinfo  {journal} {Phys. Rev. B}\ }\textbf {\bibinfo {volume} {65}},\ \bibinfo {pages} {035109} (\bibinfo {year} {2001})}\BibitemShut {NoStop}%
\bibitem [{\citenamefont {Mostofi}\ \emph {et~al.}(2014)\citenamefont {Mostofi}, \citenamefont {Yates}, \citenamefont {Pizzi}, \citenamefont {Lee}, \citenamefont {Souza}, \citenamefont {Vanderbilt},\ and\ \citenamefont {Marzari}}]{Mostofi2014An-CPC}%
  \BibitemOpen
  \bibfield  {author} {\bibinfo {author} {\bibfnamefont {A.~A.}\ \bibnamefont {Mostofi}}, \bibinfo {author} {\bibfnamefont {J.~R.}\ \bibnamefont {Yates}}, \bibinfo {author} {\bibfnamefont {G.}~\bibnamefont {Pizzi}}, \bibinfo {author} {\bibfnamefont {Y.-S.}\ \bibnamefont {Lee}}, \bibinfo {author} {\bibfnamefont {I.}~\bibnamefont {Souza}}, \bibinfo {author} {\bibfnamefont {D.}~\bibnamefont {Vanderbilt}},\ and\ \bibinfo {author} {\bibfnamefont {N.}~\bibnamefont {Marzari}},\ }\bibfield  {title} {\bibinfo {title} {An updated version of wannier90: A tool for obtaining maximally-localised wannier functions},\ }\href@noop {} {\bibfield  {journal} {\bibinfo  {journal} {Comput. Phys. Commun.}\ }\textbf {\bibinfo {volume} {185}},\ \bibinfo {pages} {2309} (\bibinfo {year} {2014})}\BibitemShut {NoStop}%
\bibitem [{\citenamefont {Lopez~Sancho}\ \emph {et~al.}(1984)\citenamefont {Lopez~Sancho}, \citenamefont {Lopez~Sancho},\ and\ \citenamefont {Rubio}}]{Sancho1984Quick-JPF}%
  \BibitemOpen
  \bibfield  {author} {\bibinfo {author} {\bibfnamefont {M.~P.}\ \bibnamefont {Lopez~Sancho}}, \bibinfo {author} {\bibfnamefont {J.~M.}\ \bibnamefont {Lopez~Sancho}},\ and\ \bibinfo {author} {\bibfnamefont {J.}~\bibnamefont {Rubio}},\ }\bibfield  {title} {\bibinfo {title} {Quick iterative scheme for the calculation of transfer matrices: application to {Mo} (100)},\ }\href@noop {} {\bibfield  {journal} {\bibinfo  {journal} {J. Phys. F: Met. Phys.}\ }\textbf {\bibinfo {volume} {14}},\ \bibinfo {pages} {1205} (\bibinfo {year} {1984})}\BibitemShut {NoStop}%
\bibitem [{\citenamefont {Lopez~Sancho}\ \emph {et~al.}(1985)\citenamefont {Lopez~Sancho}, \citenamefont {Lopez~Sancho}, \citenamefont {Sancho},\ and\ \citenamefont {Rubio}}]{Sancho1985Highly-JPF}%
  \BibitemOpen
  \bibfield  {author} {\bibinfo {author} {\bibfnamefont {M.~P.}\ \bibnamefont {Lopez~Sancho}}, \bibinfo {author} {\bibfnamefont {J.~M.}\ \bibnamefont {Lopez~Sancho}}, \bibinfo {author} {\bibfnamefont {J.~M.~L.}\ \bibnamefont {Sancho}},\ and\ \bibinfo {author} {\bibfnamefont {J.}~\bibnamefont {Rubio}},\ }\bibfield  {title} {\bibinfo {title} {Highly convergent schemes for the calculation of bulk and surface {Green} functions},\ }\href@noop {} {\bibfield  {journal} {\bibinfo  {journal} {J. Phys. F: Met. Phys.}\ }\textbf {\bibinfo {volume} {15}},\ \bibinfo {pages} {851} (\bibinfo {year} {1985})}\BibitemShut {NoStop}%
\bibitem [{\citenamefont {Wu}\ \emph {et~al.}(2018{\natexlab{b}})\citenamefont {Wu}, \citenamefont {Zhang}, \citenamefont {Song}, \citenamefont {Troyer},\ and\ \citenamefont {Soluyanov}}]{Wu2018WannierTools-CPC}%
  \BibitemOpen
  \bibfield  {author} {\bibinfo {author} {\bibfnamefont {Q.}~\bibnamefont {Wu}}, \bibinfo {author} {\bibfnamefont {S.}~\bibnamefont {Zhang}}, \bibinfo {author} {\bibfnamefont {H.-F.}\ \bibnamefont {Song}}, \bibinfo {author} {\bibfnamefont {M.}~\bibnamefont {Troyer}},\ and\ \bibinfo {author} {\bibfnamefont {A.~A.}\ \bibnamefont {Soluyanov}},\ }\bibfield  {title} {\bibinfo {title} {{WannierTools}: An open-source software package for novel topological materials},\ }\href@noop {} {\bibfield  {journal} {\bibinfo  {journal} {Comput. Phys. Commun.}\ }\textbf {\bibinfo {volume} {224}},\ \bibinfo {pages} {405} (\bibinfo {year} {2018}{\natexlab{b}})}\BibitemShut {NoStop}%
\end{thebibliography}


\end{document}